\documentclass[apj]{emulateapj}       
\usepackage{apjfonts,epsfig,natbib}   

\newcommand{\hi}{\hbox{\ion{H}{1}}}         
\newcommand{\hii}{\hbox{\ion{H}{2}}}        

\newcommand{\otwo}{\hbox{[\ion{O}{2}]$\lambda 3727$}}
\newcommand{\nethree}{\hbox{[\ion{Ne}{3}]$\lambda 3869$}}
\newcommand{\hdelta}{\hbox{H$\delta$}}
\newcommand{\hgamma}{\hbox{H$\gamma$}}
\newcommand{\othreea}{\hbox{[\ion{O}{3}]$\lambda 4363$}}
\newcommand{\hbeta}{\hbox{H$\beta$}}
\newcommand{\othree}{\hbox{[\ion{O}{3}]$\lambda\lambda 4959,5007$}}

\newcommand{\othreec}{\hbox{[\ion{O}{3}]$\lambda 5007$}}

\newcommand{\halpha}{\hbox{H$\alpha$}}

\newcommand{\stwo}{\hbox{[\ion{S}{2}]$\lambda\lambda 6716,6731$}}

\newcommand{\ntwootwo}{\hbox{[\ion{N}{2}]/[\ion{O}{2}]}}

\journalinfo{ApJ, accepted} 

\shortauthors{Lee et al.}
\shorttitle{Nebular and Stellar Abundances in WLM}
\slugcomment{{\sc Submitted: 16 Jul 2004; Version: 26 Oct 2004}}

\begin{document}

\title{
Investigating The Possible Anomaly Between Nebular and Stellar \\
Oxygen Abundances in the Dwarf Irregular Galaxy 
WLM$\,$\altaffilmark{\dag}
}

\author{
Henry Lee$\,$\altaffilmark{1}, 
Evan D. Skillman$\,$\altaffilmark{1}, and
Kim A. Venn$\,$\altaffilmark{2,1}
}

\altaffiltext{\dag}{
Based on EFOSC2 observations collected at the European Southern
Observatory, Chile: proposal \#71.D-0491(B). 
}
\altaffiltext{1}{
Department of Astronomy, University of Minnesota,
116 Church St. SE, Minneapolis, MN 55455;
{\tt hlee@astro.umn.edu, skillman@astro.umn.edu}
}
\altaffiltext{2}{
Department of Physics \& Astronomy, Macalester College,
1600 Grand Avenue, Saint Paul, MN 55105;
{\tt venn@macalester.edu}
}

\begin{abstract}				
We obtained new optical spectra of 13 \hii\ regions in WLM with EFOSC2;
oxygen abundances are derived for nine \hii\ regions.
The temperature-sensitive \othreea\ emission line was measured
in two bright \hii\ regions HM~7 and HM~9. 
The direct oxygen abundances for HM~7 and HM~9 are
12$+$log(O/H) = $7.72 \pm 0.04$ and $7.91 \pm 0.04$, respectively.
We adopt a mean oxygen abundance of 12$+$log(O/H) = $7.83 \pm 0.06$.
This corresponds to [O/H] = $-0.83$~dex, or 15\% of the solar value.
In \hii\ regions where \othreea\ was not measured, oxygen abundances
derived with bright-line methods are in general agreement with direct
values of the oxygen abundance to an accuracy of about 0.2~dex.
In general, the present measurements show that the \hii\ region oxygen
abundances agree with previous values in the literature.
The nebular oxygen abundances are marginally consistent with the mean
stellar magnesium abundance ([Mg/H] = $-0.62$).
However, there is still a 0.62~dex discrepancy in oxygen abundance 
between the nebular result and the A-type supergiant star WLM~15
([O/H] = $-0.21$).
Non-zero reddening values derived from Balmer line ratios
were found in \hii\ regions near a second \hi\ peak.
There may be a connection between the location of the second \hi\ peak, 
regions of higher extinction, and the position of WLM~15 on the
eastern side of the galaxy.
\end{abstract}				


\section{Introduction}

Dwarf galaxies are thought to be the building blocks in the assembly
of more massive galaxies within the hierarchical picture of structure
formation.
These galaxies are also very important venues in which questions about
cosmology, galaxy evolution, and star formation may be answered.
Dwarf irregular galaxies are relatively low-mass, gas-rich,
metal-poor, and are presently forming stars as shown by their \hii\
regions, whereas low-mass dwarf spheroidal galaxies are gas-poor and
no longer host present-day star-forming events.
The properties of these galaxies may be similar to those found in the
early universe, and dwarf irregulars may possibly be sites out of
which damped Lyman-$\alpha$ absorber systems form at high redshift
(e.g., \citealp{calura03,prochaska03}).
An important question which has yet to be fully explained is the
relationship between dwarf irregular and dwarf spheroidal galaxies
(e.g., \citealp{ggh03,scm03,evan_ic1613,vanzee04}, and references
therein).
That streams have been observed within the Galaxy and M~31 
(e.g., \citealp{yanny03,martin04,zucker04a,zucker04b})
has been taken as evidence of ongoing accretion and of representing
past merging of dwarfs by the more massive galaxies.
However, work presented by \cite{tolstoy03} and \cite{venn_dsph} have
shown that stars in present-day dwarf spheroidals cannot make up the
dominant stellar populations in the halo, bulge, or the thick disk of
the Galaxy, although the merging of dwarf galaxies at very early times
cannot be ruled out.

The measurements of element abundances provide important clues
to understanding the chemical history and evolution of galaxies.
In star-forming dwarf galaxies, the analysis of bright nebular
emission lines from the spectra of \hii\ regions is used to derive
abundances of $\alpha$-elements (i.e., oxygen) in the ionized
gas (see e.g., \citealp{dinerstein90,skillman98,garnett04}).
However, a limited number of elements can be studied by comparison 
to the number of elements found in the absorption spectra of stars.
For a more complete picture, additional elements should be included,
since various elements arise from different sites and involve
different timescales.
Oxygen and other $\alpha$-elements are created in very massive
progenitor stars before being returned to the interstellar medium
(ISM) on short timescales, when these stars explode as Type II
supernovae. 
Iron is an element produced by explosive nucleosynthesis in Type~I
supernovae from low-mass progenitor stars on longer timescales, and is
also produced in Type~II supernovae. 
Because of the varying timescales for stars of different masses, the
$\alpha$ element-to-iron abundance ratio, [$\alpha$/Fe],\footnote{
We use the notation: [X/Y] = log(X/Y) $-$ log(X/Y)$_{\odot}$.
} 
is tied very strongly with the star formation history 
(e.g., \citealp{gw91,matteucci03}). 
Interestingly, [$\alpha$/Fe] values for three dwarf irregular galaxies
are near or at solar, which indicates that stars have been forming at
a very low rate and/or the last burst of star formation occurred long
ago \citep{venn01,venn03,kaufer04}.
\cite{it99} claim that O/Fe is elevated in low metallicity blue
compact dwarf galaxies ([O/Fe] = $+0.32 \pm 0.11$). 
However, their analysis does not account for potential depletion of Fe
onto dust grains, and the Fe abundance is only measured in Fe$^{+2}$,
requiring very large and uncertain ionization correction factors (ICFs).
\cite{rodriguez03} finds that the adopted ICFs underestimate the total
Fe abundance by factors larger than the elevated abundance ratio
claimed by \cite{it99}.
Thus, it is prudent to assume that the nebular Fe abundances in these
galaxies, and thus the nebular O/Fe ratios, are quite uncertain
\citep{garnett04}.
At present, reliable O/Fe ratios will need to be obtained from 
stellar abundances.
While a complete discussion of $\alpha$/Fe values is
beyond the scope of the present work, brief reviews of stellar
abundances in external galaxies have recently been presented by
\cite{tv04} and \cite{venn_ociw}.

High efficiency spectrographs on 8- and 10-m telescopes have made
possible the spectroscopic measurements of individual stars in
extragalactic systems.
In particular, bright blue supergiants have been observed in galaxies
at distances of about 1~Mpc.
These hot young massive stars allow us to measure simultaneously
present-day $\alpha$- and iron-group elements.
The important advantage of these measurements also allow for the
direct comparison of stellar $\alpha$-element abundances with nebular
measurements, as massive stars and nebulae are similar in age and have
similar formation sites.
Oxygen abundances derived from the spectroscopy of blue supergiants
have been obtained in nearby dwarf irregular galaxies NGC~6822,
WLM, and Sextans~A \citep{venn01,venn03,kaufer04}.

The relative ease with which spectra of \hii\ regions have been obtained
in dwarf irregular galaxies has led to establishing:
(1) the metallicity-luminosity relation, thought to be representative
of a mass-metallicity relation for dwarf irregular galaxies
(e.g., \citealp{skh89,rm95,lee03field});
and
(2) the metallicity-gas fraction relation,
which represents the relative conversion of gas into stars, and may be
strongly affected by the galaxies' surrounding environment
(e.g., \citealp{lee03field,lee03virgo,scm03}).
It is assumed that nebular oxygen abundances are representative of the
present-day ISM metallicity for an entire dwarf galaxy, where there is
often only a single \hii\ region present.
In fact, spatial inhomogeneities or radial gradients in oxygen
abundances have been found to be very small or negligible in nearby
dwarf irregular galaxies (e.g., \citealp{ks96,ks97}), although recent
observations have cast uncertainty about the assumption in NGC~6822 and
WLM \citep{venn01,venn03}.
Here we will focus on oxygen abundances, and the comparison between
stellar and nebular determinations.
For the remainder of this paper, we adopt 12$+$log(O/H) = 8.66 as the 
solar value for the oxygen abundance
\citep{asplund04}.

\subsection{WLM}

WLM (Wolf-Lundmark-Melotte) is a dwarf irregular galaxy at a distance
of 0.95~Mpc \citep{dolphin00} and is located in the Local Group.
The galaxy was discovered by \cite{wolf10}\footnote{
On 1909 October 15, Wolf observed the galaxy in two hours with a
Waltz reflector at the Heidelberg Observatory atop K\"onigstuhl.
He submitted a short description of his observations with the 
title ``\"Uber einen gr\"osseren Nebelfleck in Cetus'' 
(On a larger hazy spot in Cetus) to Astronomische Nachrichten 
on 1909 November 16.  
}, 
and independently rediscovered by Lundmark and Melotte 
\citep{melotte26}.
WLM is relatively isolated, as the nearest neighbor about 175~kpc
distant is the recently discovered Cetus dwarf spheroidal galaxy 
\citep{whiting99}.
Basic properties of the galaxy are listed in Table~\ref{table_wlm}.

A number of observations are summarized here.
\cite{jl81} identified two planetary nebulae in the galaxy, and
\cite{sc85} identified the brightest blue and red supergiant
stars, including over 30 variable stars.
Ground-based optical photometry of stars were obtained by
\cite{ferraro89}, and \cite{mz96,mz97}.
The presence of a single globular cluster was established,
and \cite{hodge99} showed that the properties of the globular cluster
are similar to those of Galactic globular clusters.
In independent \halpha\ imaging programs, \cite{hhg93} detected two
small shell-like features, and \cite{hm95} cataloged and measured
\halpha\ fluxes for 21 \hii\ regions in the galaxy.
\cite{tomita98} presented \halpha\ velocity fields for the brightest
\hii\ regions in WLM, and showed that the southern \hii\ ring is
expanding at a speed of 20 km~s$^{-1}$ and that the kinetic age
of the bubble is 4.5~Myr.
Recent studies of the resolved stellar populations with
the Hubble Space Telescope ({\em HST\/}) have been carried out
by \cite{dolphin00} with the Wide Field Planetary Camera~2 
and by \cite{rejkuba00} with the Space Telescope Imaging Spectrograph.
\cite{dolphin00} found that over half of the stars were formed 
about 9~Gyr ago, and that a recent burst of star formation has mostly
occurred in the central bar of the galaxy.
\cite{rejkuba00} identified the horizontal branch, also confirming
the presence of a very old stellar population.
In the carbon star survey by \cite{bd04}, they found that WLM
contained the largest fraction of carbon-to-M stars for the dwarf
galaxies surveyed, and showed that WLM is an inclined disk galaxy with
no evidence of an extended spherical stellar halo.
\cite{tk01} searched for molecular gas in WLM, but only upper limits
to the CO intensity and subsequent H$_2$ column densities were 
determined.
Recent 21-cm measurements with the Australia Telescope Compact Array
have shown that there are two peaks in the \hi\ distribution, and that
the measured \hi\ rotation curve is typical for a disk
\citep{jackson04}.

The spectroscopy of the brightest \hii\ regions were reported by
\cite{stm89} and \cite{hm95}. 
The resulting nebular oxygen abundances were found to be 
12$+$log(O/H) $\simeq 7.74$, or [O/H] $\simeq -0.92$.
\cite{venn03} measured the chemical composition of two
A-type supergiant stars in WLM, and showed that the mean stellar
magnesium abundance was [Mg/H] = $-0.62$.
However, the oxygen abundance in one of the stars was [O/H] = $-0.21$,
which is about 0.7 dex or almost five times larger than the nebular
abundance.
This presents a vexing question: how can the young supergiant be
significantly more metal-rich than the surrounding ISM from which the
star was born?  

The research reported here is part of a program to understand the
chemical evolution from the youngest stellar populations in the
nearest dwarf irregular galaxies (e.g., \citealp{venn01,venn03,kaufer04}).
The motivations are:
(1) to obtain a homogeneous sample of abundance measurements for \hii\
regions presently known in WLM;
(2) to measure the temperature-sensitive \othreea\ emission line,
derive direct oxygen abundances, and compare the present set of
measurements with those in the literature; and
(3) to examine whether the present measurements show any
inhomogeneities in oxygen abundances across the galaxy.
This is the first of two papers of our study; the measurements and
analyses for \hii\ regions in NGC~6822 will be discussed in the next
paper (H. Lee et al., in preparation).
The outline of this paper is as follows.
Observations and reductions of the data are presented in
Sect.~\ref{sec_obs}.
Measurements and analyses are discussed in Sect.~\ref{sec_analysis},
and nebular abundances are presented in Sect.~\ref{sec_abund}.
Our results are discussed in Sect.~\ref{sec_discuss}, and 
a summary is given in Sect.~\ref{sec_concl}.

\section{Observations and Reductions}
\label{sec_obs}		

Long-slit spectroscopic observations of \hii\ regions in WLM were
carried out on 2003 Aug. 26--28 and 31 (UT) with the 
ESO Faint Object Spectrograph and Camera (EFOSC2) instrument on the
3.6-m telescope at ESO La Silla Observatory.
Details of the instrumentation employed and the log of observations
are listed in Tables~\ref{table_obsprops} and \ref{table_obslog},
respectively.
Observing conditions were obtained during new moon phase.
Conditions varied from photometric (26 Aug UT) to cloudy (31 Aug UT).
Two-minute \halpha\ acquisition images were obtained in
order to set an optimal position angle of the slit, so that 
the slit could cover as many \hii\ regions possible.
Thirteen \hii\ regions for which spectra were obtained are listed
in Table~\ref{table_obslog} and shown in Fig.~\ref{fig_h2rs}.
Identifications for the \hii\ regions follow from the \halpha\ imaging
compiled by \cite{hm95}\footnote{
We can also compare the locations of \hii\ regions HM~2, HM~8, and
HM~9 in the \halpha\ image by \cite{hm95} with the \othreec\ image 
from \cite{jl81}.
}.
For completeness, we provide here coordinates (Epoch J2000) for
\hii\ regions which were ``newly'' resolved in images obtained
by the Local Group Survey\footnote{
A description and distribution of the data from the Local Group Survey 
may be found at {\tt http://www.lowell.edu/users/massey/lgsurvey.html} 
\citep{massey_lgs}.
}.
HM~16 was resolved into two separate \hii\ regions, which
we have called HM~16 NW ($\alpha =$ 00$^{\rm h}$01$^{\rm m}$59\fs4; 
$\delta = -15\degr27\arcmin24\farcs9$),
and HM~16 SE ($\alpha = $ 00$^{\rm h}$01$^{\rm m}$59\fs6; 
$\delta = -15\degr27\arcmin29\farcs1$).
To the east of \hii\ region HM~18, we took spectra of two 
additional compact \hii\ regions: 
HM~18a ($\alpha = $ 00$^{\rm h}$01$^{\rm m}$59\fs7;
$\delta = -15\degr29\arcmin30\farcs1$)
and HM~18b ($\alpha = $ 00$^{\rm h}$01$^{\rm m}$59\fs9;
$\delta = -15\degr29\arcmin45\farcs5$).

Data reductions were carried out in the standard manner using 
IRAF\footnote{
IRAF is distributed by the National Optical Astronomical
Observatories, which are operated by the Associated Universities for
Research in Astronomy, Inc., under cooperative agreement with the
National Science Foundation.}
routines.
Data obtained for a given night were reduced independently.
The raw two-dimensional images were trimmed and the bias level was
subtracted.
Dome flat exposures were used to remove pixel-to-pixel variations 
in response. 
Twilight flats were acquired at dusk each night to correct
for variations over larger spatial scales.
To correct for the ``slit function'' in the spatial direction, the
variation of illumination along the slit was taken into account
using dome and twilight flats. 
Cosmic rays were removed in the addition of multiple exposures
for a given \hii\ region.
Wavelength calibration was obtained using helium-argon (He-Ar) arc
lamp exposures taken throughout each night.
Exposures of standard stars Feige~110, G138$-$31, LTT~1788, LTT~7379,
and LTT~9491 were used for flux calibration.
The flux accuracy is listed in Table~\ref{table_obslog}. 
Final one-dimensional spectra for each \hii\ region were obtained via
unweighted summed extractions.  

\section{Measurements and Analysis}
\label{sec_analysis}

Emission-line strengths were measured using software developed 
by M. L. McCall and L. Mundy; see \cite{lee01,lee03field,lee03virgo}.
\othreea\ was detected in \hii\ regions HM~7 and HM~9; these spectra
are shown in Fig.~\ref{fig_specall}.

Corrections for reddening and for underlying absorption 
and abundance analyses were performed with SNAP 
(Spreadsheet Nebular Analysis Package, \citealp{snap97}). 
Balmer fluxes were first corrected for underlying Balmer absorption
with an equivalent width 2~\AA\ \citep{mrs85,lee03virgo}.
\halpha\ and \hbeta\ fluxes were used to derive
reddening values, $E(B-V$), using the equation
\begin{equation}
\log\frac{I(\lambda)}{I(\hbeta)} = 
\log\frac{F(\lambda)}{F(\hbeta)} + 
0.4\,E(B-V)\,\left[A_1(\lambda) - A_1(\hbeta)\right]
\label{eqn_corrthesis}
\end{equation}
\citep{lee03field}.
$F$ and $I$ are the observed flux and corrected intensity ratios,
respectively.
Intrinsic case-B Balmer line ratios determined by \cite{sh95} were
assumed.
$A_1(\lambda)$ is the extinction in magnitudes for $E(B-V) = 1$, 
i.e., $A_1(\lambda) = A(\lambda)/E(B-V)$, where 
$A(\lambda)$ is the monochromatic extinction in magnitudes.
Values of $A_1$ were obtained from the \cite{cardelli89}
reddening law as defined by a ratio of the total to selective
extinction, $R_V$ = $A_V/E(B-V)$ = 3.07, which in the limit of zero
reddening is the value for an A0V star (e.g., Vega) with intrinsic
color $(B-V)^0 = 0$.
Because \stwo\ lines were generally unresolved, 
$n_e$ = 100~cm$^{-3}$ was adopted for the electron density.
Errors in the derived $E(B-V)$ were computed from the maximum and
minimum values of the reddening based upon $2\sigma$ errors in the
fits to emission lines.

Observed flux $(F)$ and corrected intensity $(I)$ ratios are listed
in Tables~\ref{table_data1} to \ref{table_data4} inclusive.
The listed errors for the observed flux ratios at each wavelength
$\lambda$ account for the errors in the fits to the line profiles,
their surrounding continua, and the relative error in the sensitivity
function stated in Table~\ref{table_obslog}.  
Errors in the corrected intensity ratios account for maximum and
minimum errors in the flux of the specified line and of 
the \hbeta\ reference line.
At the \hbeta\ reference line, errors for both observed and corrected
ratios do not include the error in the flux.
Also given for each \hii\ region are: the observed \hbeta\ flux, the
equivalent width of the \hbeta\ line in emission, and the derived 
reddening from SNAP.

Where \othreea\ is measured, we also have performed the additional 
computations to check the consistency of our results.
Equation~(\ref{eqn_corrthesis}) can be generalized and rewritten as
\begin{equation}
\log\frac{I(\lambda)}{I(\hbeta)} =
\log\frac{F(\lambda)}{F(\hbeta)} + c(\hbeta) \, f(\lambda),
\label{eqn_corr}
\end{equation}
where $c(\hbeta)$ is the logarithmic extinction at \hbeta, and 
$f(\lambda)$ is the wavelength-dependent reddening function 
\citep{aller84,osterbrock}.
From Equations~(\ref{eqn_corrthesis}) and (\ref{eqn_corr}), 
we obtain
\begin{equation}
c(\hbeta) = 1.43 \, E(B-V) = 0.47 \, A_V.
\label{eqn_chbebv}
\end{equation}
The reddening function normalized to \hbeta\ is derived from
the \cite{cardelli89} reddening law, assuming $R_V$ = 3.07.
As described in \cite{scm03}, values of $c(\hbeta)$
were derived from the error weighted average of values for
$F(\halpha)/F(\hbeta)$, $F(\hgamma)/F(\hbeta)$, and
$F(\hdelta)/F(\hbeta)$ ratios, while simultaneously solving for the
effects of underlying Balmer absorption with equivalent width 
EW$_{\rm abs}$.
We assumed that EW$_{\rm abs}$ was the same for \halpha, \hbeta,
\hgamma, and \hdelta.
Uncertainties in $c(\hbeta)$ and EW$_{\rm abs}$ were determined
from Monte Carlo simulations \citep{os01,scm03}.
Errors derived from these simulations are larger than errors
quoted in the literature by either assuming a constant value for
the underlying absorption or derived from a $\chi^2$ analysis in
the absence of Monte Carlo simulations for the errors; 
Fig.~\ref{fig_monte} shows an example of these simulations for \hii\
region HM~9.
In Tables~\ref{table_data1} to \ref{table_data3}, we included
the logarithmic reddening and the equivalent width of the underlying
Balmer absorption, which were solved simultaneously.
Values for the logarithmic reddening are consistent with values
of the reddening determined with SNAP.
Where negative values were derived, the reddening was set to zero
in correcting line ratios and in abundance calculations.

\section{Nebular Abundances}
\label{sec_abund}

Oxygen abundances in \hii\ regions were derived using three methods:
(1) the direct method 
(e.g., \citealp{dinerstein90,skillman98,garnett04});
and two bright-line methods discussed by
(2) \cite{mcgaugh91}, which is based on photoionization models;  
and (3) \cite{pilyugin00}, which is purely empirical.

\subsection{Oxygen Abundances: \othreea\ Temperatures}

For the ``direct'' conversion of emission-line intensities into ionic
abundances, a reliable estimate of the electron temperature in the
ionized gas is required.
To describe the ionization structure of \hii\ regions, we adopt
a two-zone model, with a low- and a high-ionization zone characterized
by temperatures $T_e($O$^+)$ and $T_e($O$^{+2})$, respectively. 
The temperature in the O$^{+2}$ zone is measured with
the emission-line ratio $I$(\othreec)/$I$(\othreea) 
\citep{osterbrock}.
The temperature in the O$^+$ zone is given by 
\begin{equation}
t_e([{\rm O\;II}]) = 0.7 \, t_e([{\rm O\;III}]) + 0.3,
\label{eqn_toplus}
\end{equation}
where $t_e = T_e/10^4$~K \citep{ctm86,garnett92}.
The uncertainty in $T_e($O$^{+2})$ is computed from the maximum
and minimum values derived from the uncertainties in corrected 
emission line ratios.
The computation does not include uncertainties in the reddening (if any),
the uncertainties in the atomic data, or the presence of temperature
fluctuations.
The uncertainty in $T_e$(O$^+$) is assumed to be the same
as the uncertainty in $T_e$(O$^{+2}$).
These temperature uncertainties are conservative estimates, and are
likely overestimates of the actual uncertainties.
For subsequent calculations of ionic abundances, we assume
the following electron temperatures for specific ions
\citep{garnett92}: 
$t_e$(N$^+$) = $t_e$(O$^+$), 
$t_e$(Ne$^{+2}$) = $t_e$(O$^{+2}$), 
$t_e$(Ar$^{+2}$) = 0.83 $t_e$(O$^{+2}$) $+$ 0.17, and
$t_e$(Ar$^{+3}$) = $t_e$(O$^{+2}$).

The total oxygen abundance by number is given by
O/H = O$^0$/H $+$ O$^+$/H $+$ O$^{+2}$/H $+$ O$^{+3}$/H.
For conditions found in typical \hii\ regions and those presented
here, very little oxygen in the form of O$^0$ is expected, and is not
included here.
In the absence of He~II emission, the O$^{+3}$ contribution is
considered to be negligible.
Ionic abundances for O$^+$/H and O$^{+2}$/H were computed using
O$^+$ and O$^{+2}$ temperatures, respectively, as described above.

Measurements of the \othreea\ line were obtained and subsequent
electron temperatures were derived in HM~7 and HM~9.
Ionic abundances and total abundances are computed using the 
method described by \cite{lee03field}. 
With SNAP, oxygen abundances were derived using the five-level atom
approximation \citep{fivel}, and transition probabilities and collision
strengths for oxygen from \cite{pradhan76}, \cite{mb93}, \cite{lb94}, 
and \cite{wiese96}; Balmer line emissivities from \cite{sh95} were used.
Derived ionic and total abundances are listed in
Tables~\ref{table_abund1} and \ref{table_abund2}.
These tables include derived O$^+$ and O$^{+2}$ electron temperatures, 
O$^+$ and O$^{+2}$ ionic abundances, and the total oxygen abundances.
Errors in direct oxygen abundances computed with SNAP have two
contributions: the maximum and minimum values for abundances from
errors in the temperature, and the maximum and minimum possible values
for the abundances from propagated errors in the intensity ratios. 
These uncertainties in oxygen abundances are also conservative
estimates. 

Using the method described by \cite{scm03}, we recompute oxygen
abundances in \hii\ regions with \othreea\ detections.
Abundances are computed using the emissivities from the five-level
atom program by \cite{sd95}.
As described above, we use the same two-temperature zone model and 
temperatures for the remaining ions.
The error in $T_e$(O$^{+2}$) is derived from the uncertainties in
the corrected emission-line ratios, and does not include any
uncertainties in the atomic data, or the possibility of temperature
variations within the O$^{+2}$ zone.
The fractional error in $T_e$(O$^{+2}$) is applied similarly to
$T_e$(O$^+$) to compute the uncertainty in the latter.
Uncertainties in the resulting ionic abundances are combined in
quadrature for the final uncertainty in the total linear (summed)
abundance.
The adopted \othreea\ abundances and their uncertainties computed in
this manner are listed in Tables~\ref{table_abund1} and
\ref{table_abund2}.
Direct oxygen abundances computed with SNAP are in excellent agreement
with direct oxygen abundances computed with the method described
by \cite{scm03}; abundances from the two methods agree to within 
0.02~dex.

Direct oxygen abundances for \hii\ regions HM~7 and HM~9 are
12$+$log(O/H) = $7.72 \pm 0.04$, and $7.91 \pm 0.04$,
respectively; the latter is a weighted mean of the three 
measured values shown in Table~\ref{table_abund2}.
%
%
The mean oxygen abundance for these two \hii\ regions is 
(O/H) = $(6.69 \pm 0.93) \times 10^{-5}$, or
12$+$log(O/H) = $7.83 \pm 0.06$. 
This value corresponds to [O/H] = $-0.83$, or 15\% of the solar value.
For historical completeness, we note that our derived nebular 
oxygen abundance would correspond to [O/H] = $-1.10$ for the
\cite{ag89} value of the solar oxygen abundance. 

\subsection{Oxygen Abundances: Bright-Line Methods}
\label{sec_brightline}

For \hii\ regions without \othreea\ measurements, secondary methods
are necessary to derive oxygen abundances.
The bright-line method is so called because the oxygen abundance is
given in terms of the bright [O~II] and [O~III] lines. 
\cite{pagel79} suggested using
\begin{equation}
R_{23} = \frac{I({\otwo}) + I({\othree})}{I({\hbeta})}
\label{eqn_r23_def}
\end{equation}
as an abundance indicator.
Using photoionization models, \cite{skillman89} showed that bright
[O~II] and [O~III] line intensities can be combined to determine
uniquely the ionization parameter and an ``empirical'' oxygen
abundance in low-metallicity \hii\ regions.
\cite{mcgaugh91} developed a grid of photoionization models and
suggested using $R_{23}$ and $O_{32}$ = $I$(\othree)/$I$(\otwo) 
to estimate the oxygen abundance\footnote{
Analytical expressions for the McGaugh calibration 
can be found in \cite{chip99}.
}.
However, the calibration is degenerate such that for a given value of
$R_{23}$, two values of the oxygen abundance are possible.
The \ntwootwo\ ratio was suggested 
(e.g., \citealp{mrs85,mcgaugh94,vanzee98})
as the discriminant to choose between the ``upper branch'' (high
oxygen abundance) or the ``lower branch'' (low oxygen abundance).
In the present set of spectra, [\ion{N}{2}] line strengths are
generally small, and \ntwootwo\ has been found to be less than the
threshold value of 0.1.
\cite{pilyugin00} proposed a new calibration of the oxygen
abundances using bright oxygen lines.
At low abundances (12$+$log(O/H) $\la$ 8.2), his calibration is
expressed as 
\begin{equation}
12+\log({\rm O/H}) = 6.35 + 3.19 \log R_{23} - 1.74 \log R_3,
\label{eqn_pily}
\end{equation}
where $R_{23}$ is given by Equation~(\ref{eqn_r23_def})
and $R_3$ = $I$(\othree)/$I$(\hbeta).
In some instances, oxygen abundances with the McGaugh method could
not be computed, because the $R_{23}$ values were outside of the
effective range for the models.
\cite{scm03} have shown that the Pilyugin calibration covers
the highest values of $R_{23}$.

Oxygen abundances derived using the McGaugh and Pilyugin bright-line
calibrations are listed in Tables~\ref{table_abund1} and
\ref{table_abund2}.
For each \hii\ region, differences between direct and bright-line
abundances are shown as a function of $O_{32}$ and $R_{23}$
in Fig.~\ref{fig_oxydiff}.
The difference between the McGaugh and Pilyugin calibrations 
(indicated by asterisks) appears to correlate with log $O_{32}$, 
which has been previously noticed by \cite{scm03}, \cite{lee03south},
and \cite{ls04_n1705}.
Despite the small number of \othreea\ detections, we find 
that bright-line abundances with the McGaugh and the Pilyugin
calibrations are $\approx$ 0.10 to 0.15~dex larger and $\approx$ 0.05
to 0.10~dex smaller, respectively, than the corresponding direct
abundances.
We note also that \hii\ region HM~19 exhibits the lowest values of
$R_{23}$ and $O_{32}$ (log $R_{23}$ = 0.357, log $O_{32}$ = $-0.629$),
and is the outlier in the lower left corner of both panels in
Fig.~\ref{fig_oxydiff}.
Generally, in the absence of \othreea, an estimate of the oxygen
abundance from the bright-line calibration is good to within
$\approx$ 0.2~dex.

\subsection{Element Ratios}

We consider next argon-to-oxygen, nitrogen-to-oxygen,
and neon-to-oxygen ratios, which are listed in
Tables~\ref{table_abund1} and \ref{table_abund2}.
For metal-poor galaxies, it is assumed that 
N/O $\approx$ N$^+$/O$^+$ \citep{garnett90} and N$^+$/O$^+$ 
values were derived.
Nitrogen abundances were computed as N/H = ICF(N) $\times$ (N$^+$/H).
The ionization correction factor, ICF(N) = O/O$^+$, accounts for
missing ions. 
The resulting nitrogen-to-oxygen abundance ratios were found
to be the same as the N$^+$/O$^+$ values.
The mean value of log(N/O) = $-1.55 \pm 0.08$ is in agreement with 
the average for metal-poor blue compact dwarf galaxies \citep{it99}.

Neon abundances are derived as Ne/H = ICF(Ne) $\times$ (Ne$^{+2}$/H).
The ionization correction factor for neon is ICF(Ne) = O/O$^{+2}$.
%
%
The mean log(Ne/O) is $-0.56 \pm 0.04$, which is marginally consistent
with the range of Ne/O values at this metallicity found by \cite{it99}
and \cite{garnett04}.
However, it is 0.16 dex higher than the mean value of $-0.72 \pm 0.06$
found for blue compact galaxies by \cite{it99}\footnote{
Note that part of the difference is due to a 14\% difference in the
ratio of the \othreec\ and \nethree\ emissivities used by
\cite{it99} and that computed by the IONIC program in the NEBULAR code
of \cite{sd95}.
This 14\% difference, which translates into a 0.06 dex difference (in
the sense observed), is probably an indication of the minimum
systematic uncertainty in the atomic data which are used for
calculating nebular abundances (see \citealp{garnett04}).
}.          
Since the \nethree/\othreec\ ratio is sensitive to the reddening
correction, and our reddening corrections are only based on the
H$\alpha$/H$\beta$ ratio, we revisited this correction.
In the spectrum for \hii\ region HM~7, higher-order Balmer
lines (H9, H10, and H11) were detected (Table~\ref{table_data1}).
Their intensity ratios with respect to \hbeta\ were found to be
consistent with expected values for \hii\ regions at 
a temperature of $T_e({\rm O}^{+2})$ = 15000~K.
In the spectra (grating \#11) for \hii\ region HM~9, the closest
unblended Balmer line to \nethree\ is \hdelta, because H8 is blended
with an adjacent helium line, and H$\epsilon$ is blended with adjacent
[\ion{Ne}{3}] and helium lines.
We find that corrected \hdelta/\hbeta\ and \hgamma/\hbeta\ ratios are
consistent with expected Balmer ratios for $T_e({\rm O}^{+2})$ between
13000 and 14000~K. 

Argon is more complex, because the dominant ion is not found in
just one zone.
Ar$^{+2}$ is likely to be found in an intermediate area between the
O$^+$ and O$^{+2}$ zones.
Following the prescription by \cite{itl94}, the argon abundance,
was derived as Ar/H = ICF(Ar) $\times$ Ar$^{+2}$/H.
The ionization correction factor is given by 
ICF(Ar) = Ar/Ar$^{+2}$ = $[0.15 + x(2.39 - 2.64x)]^{-1}$,
where $x$ = O$^+$/O.
Our mean value of log(Ar/O) = $-2.12 \pm 0.12$ is in agreement with
the average for metal-poor blue compact dwarf galaxies \citep{it99}.

We note here the recent work by \cite{moore04}, who suggested that
the direct modeling of photoionized nebulae should be used
to infer elemental abundances with accuracies similar to the
observations.
They showed that abundances derived from model-based ionization
correction factors exceeded the range of expected errors from the
original data.

\section{Discussion}
\label{sec_discuss}

A comparison of the present data with published spectroscopy 
(\othreea\ detections) is shown in Table~\ref{table_compare}. 
\cite{stm89} obtained spectra of HM~9 and HM~2 (their \hii\ regions
``\#1'' and ``\#2'', respectively), while \cite{hm95} reported spectra
for \hii\ regions HM~7 and HM~9.
We have recomputed and added uncertainties to their published
abundances in Table~\ref{table_compare}.
From our measurements presented above, \othreea\ was detected in
\hii\ regions HM~7 and HM~9 at a signal-to-noise of about $10\sigma$
and $5\sigma$, respectively.
Our derived direct oxygen abundance for HM~7 is in agreement with the
value reported by \cite{hm95}.
The direct oxygen abundance for HM~9 is about 0.1~dex higher but
consistent within errors with the values reported by \cite{stm89} and
\cite{hm95}.
When other measurements of \hii\ regions are included, oxygen
abundances derived with bright-line methods agree with the
direct values to within 0.2~dex (see Fig.~\ref{fig_oxydiff}).
The present set of measurements have shown that the nebular oxygen
abundances are in agreement with values published in the literature.

Fig.~\ref{fig_oxymb} shows the metallicity-luminosity relation for
dwarf irregular galaxies.
We have followed Fig.~10 from \cite{venn03} and plotted
the present nebular result, the mean magnesium abundance from the two
A-type supergiant stars, and the derived oxygen abundance for the
supergiant WLM~15.
The upper end of the range of nebular oxygen abundances derived in the
present work 
([O/H] between $-0.98$ and $-0.71$) 
is consistent with the mean stellar magnesium abundance 
[Mg/H] = $-0.62$.
However, the mean nebular oxygen abundance is 0.62~dex lower than the
oxygen abundance ([O/H] = $-0.21$) in WLM~15.
We note that the mean stellar [Mg/H] in supergiant stars agrees with
the nebular [O/H] in both NGC~6822 \citep{venn01} and 
Sextans~A \citep{kaufer04}.
Measurements of additional supergiants in WLM will be crucial in
confirming the result in WLM~15, and whether there are any spatial
inhomogeneities in metallicity among stars.

\cite{jackson04} reported two peaks in the \hi\ distribution; the west
maximum lies just north of HM~9, whereas the second maximum lies
approximately 1\arcmin\ east of HM~7.
Figure~\ref{fig_wlm} shows that \hii\ regions along the eastern edge
of the galaxy in the vicinity of the east \hi\ peak have nonzero
reddening values derived from observed Balmer emission-line ratios.
This agrees with the discovery of small regions or patches found to
have redder $U-B$ and $B-V$ colors than the rest of the galaxy
\citep{jackson04}.
This is thought to be extinction internal to WLM, because the
foreground extinction is known to be small (see Table~\ref{table_wlm}).
We note that the bright \hii\ regions (i.e., HM~7 and HM~9) on the
western side of the galaxy do not exhibit any appreciable reddening.
A comparison of Figs.~\ref{fig_h2rs} and \ref{fig_wlm} shows that the
two supergiant stars measured by \cite{venn03} are located next to
the eastern \hi\ peak.
In fact, the supergiant WLM~15 with the high oxygen abundance is
located (in projection) near \hii\ region HM~18, where the average
reddening was found to be $E(B-V) \ga$ 0.5~mag.

We have shown that the present nebular oxygen abundance agrees with
previous measurements, and additional explanations are required to
explain the discrepancy between the nebular and stellar oxygen
abundances.
\cite{venn03} reasoned that large changes in stellar parameters
would not reconcile the stellar abundance with the nebular abundance.
Three alternative scenarios were considered to explain the
discrepancy:
lowering the nebular abundance through dilution by the infall
of metal-poor \hi\ gas, 
the depletion of ISM oxygen onto dust grains,
and the possibility of large spatial inhomogeneities in the
metallicity. 
The first two possibilities were shown to be unlikely.
The third possibility may be possible with the discovery of the second 
\hi\ peak.
However, we have shown that there are no significant variations in the
nebular oxygen abundance.

Nevertheless, two questions arise.
Are the nonzero reddenings measured in \hii\ regions along the
eastern side of the galaxy related to the second \hi\ peak seen by
\cite{jackson04}?

Because of their proximity, is there any relationship between the
the metallicity of WLM~15 and the highly reddened \hii\ regions 
in the southeast? 
We note that determining the oxygen abundance for the supergiant is
not sensitive to internal reddening \citep{venn03}.
Based on the apparent spatial correlation of the \hi\ peak
with regions of redder color, 
nonzero (internal) extinctions in \hii\ regions to the east and
southeast, and  
the location of the supergiant WLM~15,
these suggest that something unusual may be happening in this
more metal-rich part of the galaxy.
Unfortunately, \hii\ regions HM~18, 18a, and 18b are 
underluminous and exhibit very weak nebular emission; 
there are few additional bright \hii\ regions in the area.
Deeper spectroscopy of these \hii\ regions may prove illuminating.
However, measurements of additional blue supergiants in the center and
the eastern side of the galaxy (e.g., WLM~30; \citealp{venn03}) could
show whether the stellar abundances remain high with respect to
the nebular values and whether the stellar abundances are also
spatially homogeneous.
A new search for molecular gas along the east and southeast sides of
the galaxy (especially in the vicinity of WLM~15) would also be timely.
While \cite{tk01} did not detect CO, their pointings missed regions
of interest on the eastern side of the galaxy.

\section{Conclusions}		
\label{sec_concl}		

Optical spectra of 13 \hii\ regions were obtained in WLM, and oxygen
abundances were derived in nine \hii\ regions.
\othreea\ was measured in bright \hii\ regions HM~7 and HM~9.
The resulting direct oxygen abundance for HM~7 is in agreement
with previously published values.
Our $\simeq 5\sigma$ detection of \othreea\ in HM~9 confirms
the lower signal-to-noise measurements reported by 
\cite{stm89} and \cite{hm95}.
For the remaining \hii\ regions, oxygen abundances derived with
bright-line methods are accurate to about 0.2~dex.
We adopt for WLM a mean nebular oxygen abundance 
12$+$log(O/H) = $7.83 \pm 0.06$, which corresponds to 
[O/H] = $-0.83$, or 15\% of the solar value.
The upper end of the range of derived nebular oxygen abundances just
agrees with the mean stellar magnesium abundance reported
by \cite{venn03}, but the present mean nebular result is still 
0.62~dex lower than the oxygen abundance derived for the A-type
supergiant WLM~15.
Significant reddening values derived from observed Balmer
emission-line ratios were found in \hii\ regions on the eastern
side of the galaxy near one of the \hi\ peaks discovered by
\cite{jackson04}.
There may be a relationship between the location of the east \hi\
peak, regions of redder color (higher extinction), large reddenings
derived from Balmer emission-line ratios in \hii\ regions along the
eastern side of the galaxy, and the location of WLM~15.

\begin{acknowledgements}	

We thank the anonymous referee for comments which improved the
presentation of this paper.
H.~L. thanks Dale Jackson for a copy of his figures.
We are grateful to ESO for awarded telescope time, and
Lisa Germany and the staff at ESO La Silla for their help in acquiring
the spectra.
H.~L. and E.~D.~S. acknowledge partial support from a NASA LTSARP grant
NAG~5--9221 and from the University of Minnesota.
K.~A.~V. thanks the National Science Foundation for support through
a CAREER award AST~99--84073. 
For their one-year visit, E.~D.~S. and K.~A.~V. thank the Institute 
of Astronomy, University of Cambridge for their hospitality and support.
This research has made use of NASA's Astrophysics Data System, and
of the NASA/IPAC Extragalactic Database (NED), which is operated by
the Jet Propulsion Laboratory, California Institute of Technology,
under contract with the National Aeronautics and Space Administration. 

\end{acknowledgements}


\clearpage	

\begin{figure}
\figurenum{1}
\caption{
Locations in WLM of \hii\ regions for which spectra were taken.
This is an unsubtracted \halpha\ image from the Local Group Survey
\citep{massey_lgs}. 
North is at the top, and east is to the left.
Black objects on the image indicate bright sources.
The field of view shown is approximately 3\farcm2 by 3\farcm6.
Labels for \hii\ regions are from \cite{hm95}, except for two compact
sources HM~18a and HM~18b.
Also marked are two supergiant stars, WLM~15 and WLM~31, for which
spectra were measured and analyzed by \cite{venn03}.
The horizontal bar at the bottom right marks an angular scale
of 30\arcsec\ on the sky.
}
\label{fig_h2rs}
\end{figure}


\begin{figure}
\figurenum{2a}
\epsscale{0.75}
\plotone{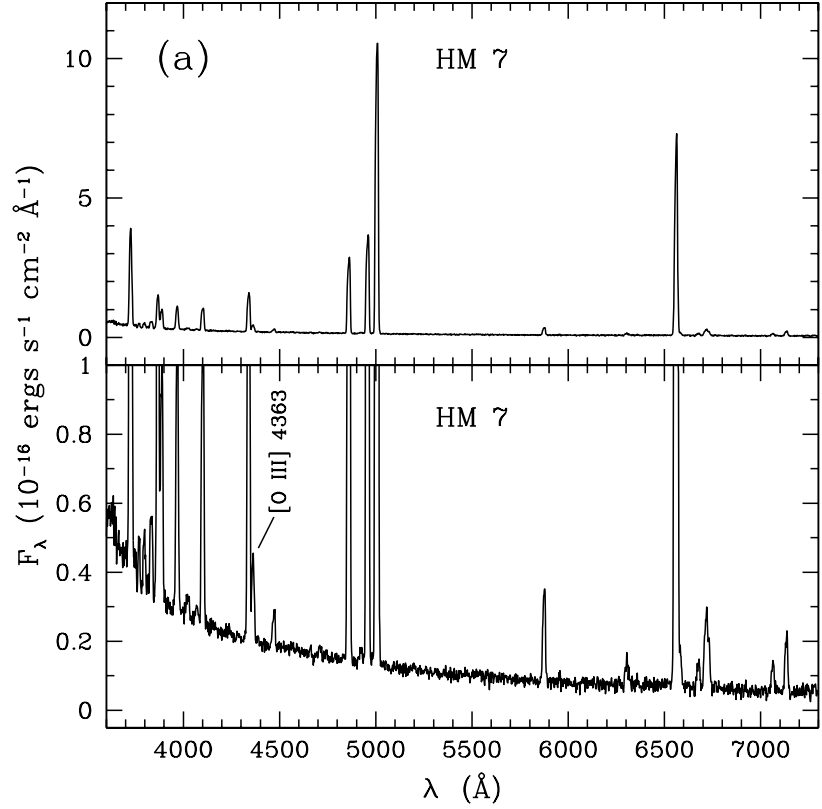}
\caption{
Emission-line spectra between 3600 and 7300~\AA\ with grating \#11.
The observed flux per unit wavelength is plotted versus wavelength.
The bottom panel is expanded to highlight \othreea.
(a) \hii\ region HM~7.
(b) \hii\ region HM~9.
}
\label{fig_specall}
\end{figure}


\begin{figure}
\figurenum{2b}
\epsscale{0.75}
\plotone{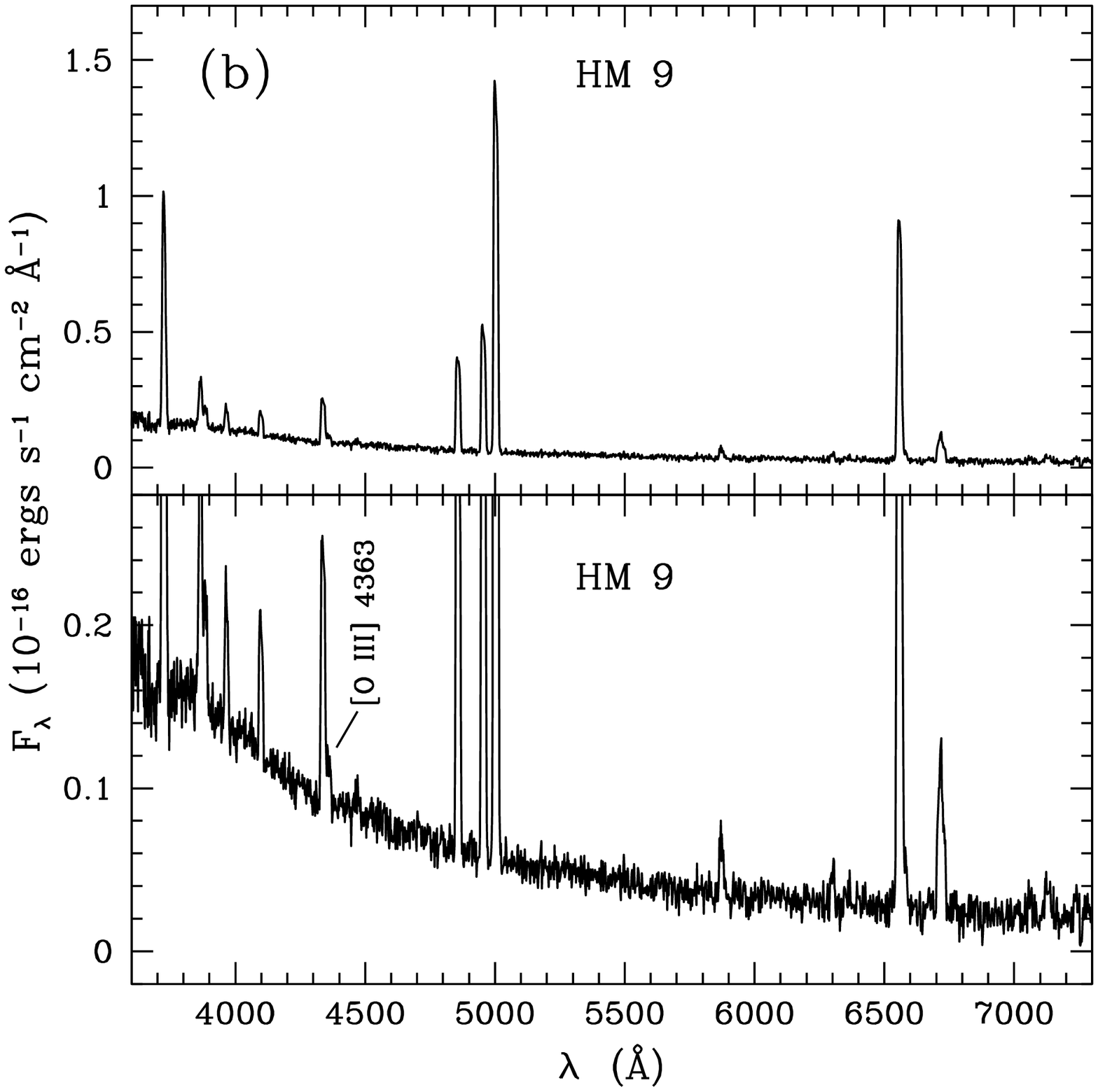}
\end{figure}


\begin{figure}
\figurenum{3}
\caption{
Monte Carlo simulations of solutions for the reddening, $c(\hbeta)$,
and the underlying Balmer absorption with equivalent width, 
EW$_{\rm abs}$, from hydrogen Balmer flux ratios.
Dotted lines mark zero values for each quantity.
The results here are shown for the \hii\ region HM~9.
Each small point is a solution derived from a different realization
of the same input spectrum.
The large filled circle with error bars shows the mean result with
$1\sigma$ errors derived from the dispersion in the solutions.
}
\label{fig_monte}
\end{figure}


\begin{figure}
\figurenum{4}
\epsscale{0.65}
\plotone{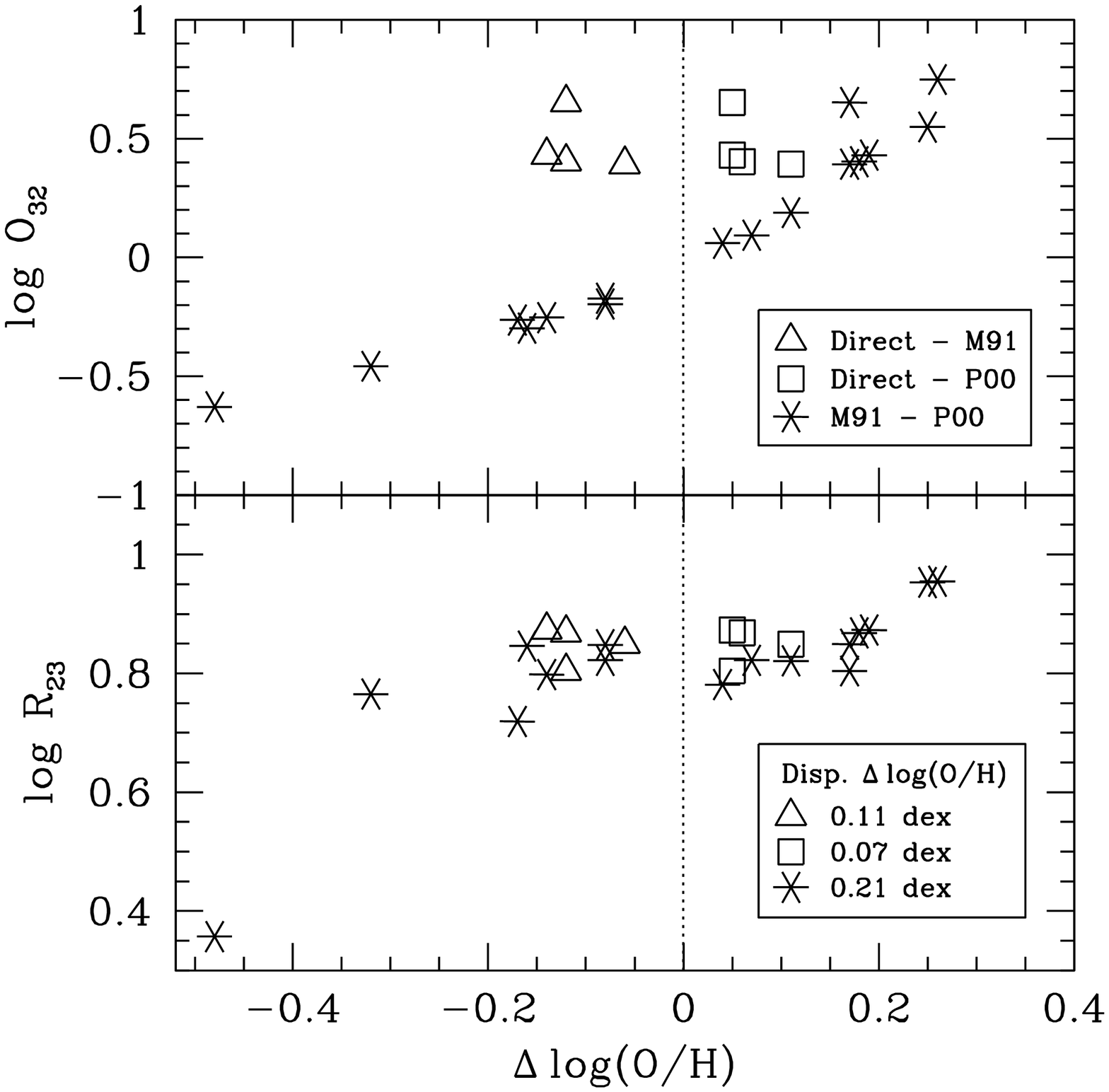}
\caption{
Difference in oxygen abundance from various methods
versus log~$O_{32}$ (top panel), and 
versus log~$R_{23}$ (bottom panel).
Each symbol represents an \hii\ region.
``Direct'' denotes oxygen abundances derived from \othreea\
measurements, ``M91'' denotes oxygen abundances derived using
the bright-line method by \cite{mcgaugh91}, and ``P00'' denotes
oxygen abundances derived using the bright-line method 
by \cite{pilyugin00}.
Vertical dotted lines in both panels mark zero differences in 
oxygen abundance.
Dispersions in abundance differences are indicated in the legend
of the bottom panel.
In the absence of \othreea, oxygen abundances derived with
the bright-line method are accurate to within $\approx$ 0.2~dex.
}
\label{fig_oxydiff}
\end{figure}


\begin{figure}
\figurenum{5}
\epsscale{1.}
\plotone{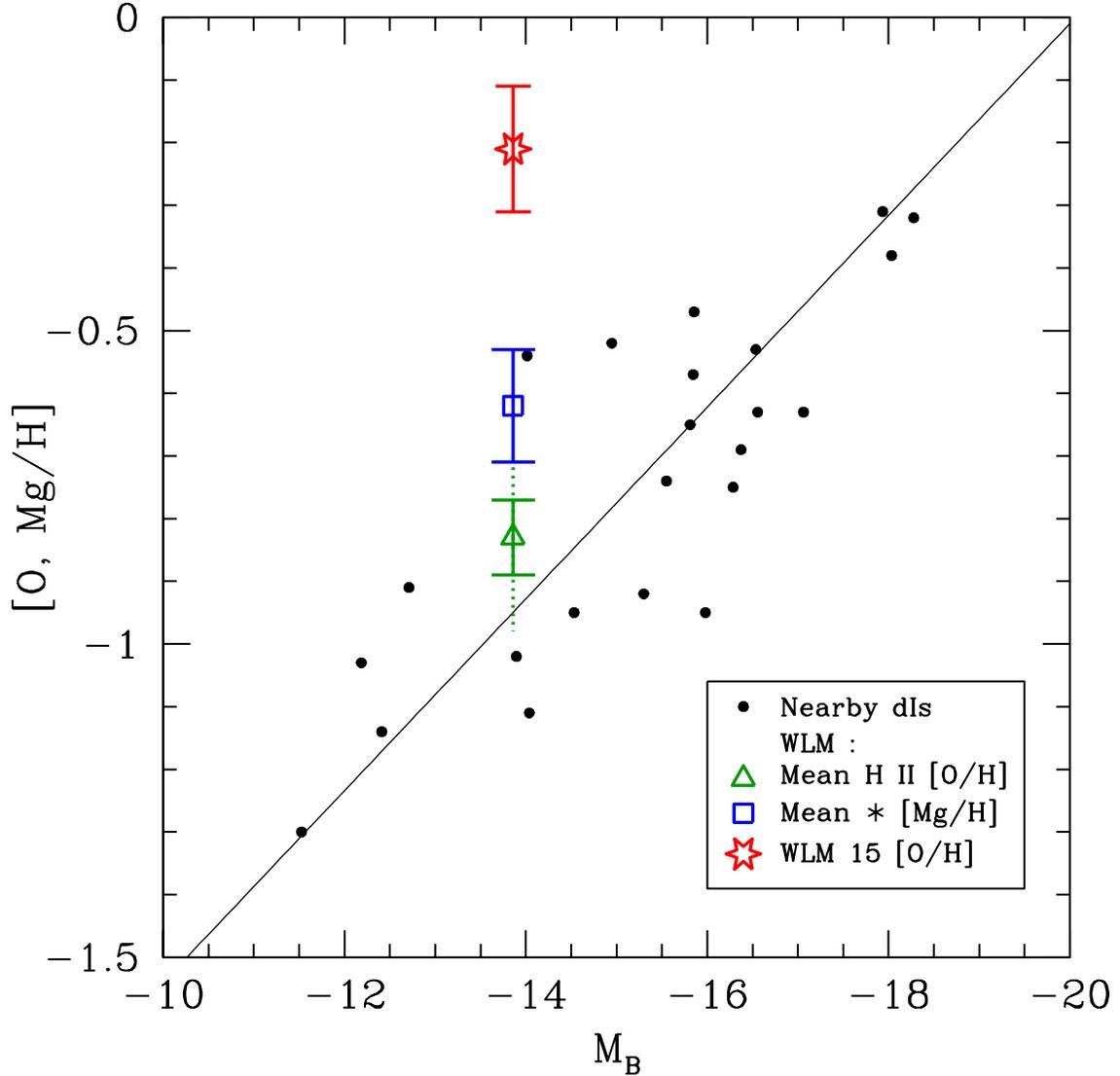}
\caption{
Metallicity-luminosity relationship for dwarf irregular galaxies.
The filled circles represent nebular oxygen abundances for the sample
of nearby dwarfs, and the solid line is the best fit
\citep{lee03south,lee03field}.
The open triangle indicates the mean nebular oxygen abundance derived
in the present work; the vertical dotted line above and below the
triangle marks the maximum ($-0.71$) and minimum ($-0.98$) 
derived values of the nebular oxygen abundance.
The open square and the open star indicate the mean [Mg/H] for the two
WLM supergiants and the stellar oxygen abundance for WLM~15,
respectively \citep[][see also their Fig.~10]{venn03}.
}
\label{fig_oxymb}
\end{figure}


\begin{figure}
\figurenum{6}
\caption{
WLM in $U$, \halpha, and \hi.  
Dark objects on the image indicate bright sources.
Left panel: \hi\ contours superposed on a $U$-band image 
(Local Group Survey, \citealp{massey_lgs}).
Right panel: \hi\ contours superposed on a continuum-subtracted
\halpha\ image (Local Group Survey, \citealp{massey_lgs}).
Open circles surround \hii\ regions where the reddening, $E(B-V)$,
was found to be non-zero.
In both panels, north is to the top and east is to the left,
and the field of view shown is 6\farcm9 by 15\farcm3.
The central rectangle in the right panel encompasses the 3\farcm2 by
3\farcm6 area shown in Fig.~\ref{fig_h2rs}. 
The figures are adapted from \cite{jackson04}.
}
\label{fig_wlm}
\end{figure}

\clearpage	

\begin{table}
\normalsize 
\tablenum{1}
\begin{center}
\renewcommand{\arraystretch}{1.}
\caption{
Basic data for WLM.
\vspace*{3mm}
\label{table_wlm}
}
\begin{tabular}{ccc}
\tableline \tableline
Property & Value & References \\
\tableline
Type & IB(s)m & \nodata \\
Alternate Names & \phn\phn DDO 221, UGCA 444 \phn\phn & \nodata \\
Distance & $0.95 \pm 0.04$ Mpc & 1 \\
Linear to angular scale at this distance & 4.6 pc arsec$^{-1}$ & 2 \\
$B_{T}\,$\tablenotemark{a} & $11.03 \pm 0.08$ & 3 \\
$E(B-V)\,$\tablenotemark{b} & 0.037 & 4 \\
$F_{21}\,$\tablenotemark{c} & $299.8 \pm 24.5$ Jy km s$^{-1}$ & 5 \\
$v_{\rm max}\,$\tablenotemark{d} & $38 \pm 5$ km s$^{-1}$ & 6 \\
$[<{\rm Mg/H}>]$\tablenotemark{e} & 
	$-0.62 \pm 0.09$ ($\pm$ 0.26)\tablenotemark{f} & 7 \\
$[{\rm O/H}]$, WLM~15$\,$\tablenotemark{g} & 
	$-0.21 \pm 0.10$ ($\pm$ 0.05)\tablenotemark{f} & 7 \\
$[<{\rm O/H}>]$, \hii\ $\,$\tablenotemark{h} & $-0.83 \pm 0.06$ & 2 \\
\tableline
\vspace*{1mm}
\end{tabular}
\tablenotetext{a}{
Apparent total $B$ magnitude.
}
\tablenotetext{b}{
Foreground reddening to the galaxy.
}
\tablenotetext{c}{
21-cm flux integral.
}
\tablenotetext{d}{
Maximum rotation velocity at the last measured point ($r$ = 0.89~kpc). 
}
\tablenotetext{e}{
Mean magnesium abundance from supergiants WLM~15 and WLM~31
with the solar value from \cite{gs98}.
}
\tablenotetext{f}{
The first uncertainty represents line-to-line scatter, and the second
uncertainty in parentheses is an estimate of the systematic error
due to certainties in stellar atmospheric parameters \citep{venn03}.
}
\tablenotetext{g}{
Oxygen abundance measured for the supergiant WLM~15 with the solar
value from \cite{asplund04}.
}
\tablenotetext{h}{
Mean \othreea\ oxygen abundance from \hii\ regions HM~7 and HM~9.
}
\tablerefs{
1.~\cite{dolphin00}; 
2.~present work;
3.~\cite{rc3};
4.~\cite{schlegel98};
5.~\cite{bdb04}; 
6.~\cite{jackson04};
7.~\cite{venn03}.
}
\end{center}
\end{table}


\begin{table}
\normalsize  
\tablenum{2}
\begin{center}
\renewcommand{\arraystretch}{1.1}
\caption{
Properties of EFOSC2 spectrograph employed at the ESO La Silla
3.6-m telescope.
\vspace*{3mm}
\label{table_obsprops}
}
\begin{tabular}{ccc}
\tableline \tableline
\multicolumn{3}{c}{{\sf Loral CCD (\#40)}} \\ 
\tableline
Total area & \multicolumn{2}{c}{2048 $\times$ 2048 pix$^2$} \\
Field of view & \multicolumn{2}{c}{5\farcm2 $\times$ 5\farcm2} \\
Pixel size & \multicolumn{2}{c}{15 $\mu$m} \\
Image scale & \multicolumn{2}{c}{0\farcs16 pixel$^{-1}$} \\
Gain & \multicolumn{2}{c}{1.3 $e^-$ ADU$^{-1}$} \\ 
Read-noise (rms) & \multicolumn{2}{c}{9 $e^-$} \\ 
\tableline
\multicolumn{3}{c}{{\sf Long slit}} \\ \tableline
Length & \multicolumn{2}{c}{$\simeq 5$\arcmin} \\
Width & \multicolumn{2}{c}{1\farcs5} \\
\tableline
& {\sf Grating \#11} & {\sf Grating \#7} \\ \tableline
Groove density & \phn 300 lines mm$^{-1}$ \phn & 600 lines mm$^{-1}$ \\
Blaze $\lambda$ (1st order) & 4000 \AA & 3800 \AA \\
Dispersion & 2.04 \AA\ pixel$^{-1}$ & 0.96 \AA\ pixel$^{-1}$ \\
Effective $\lambda$ range & 3380--7520 \AA & 3270--5240 \AA \\ 
\tableline
\end{tabular}
\end{center}
\end{table}


\begin{table}
\normalsize 
\tablenum{3}
\begin{center}
\renewcommand{\arraystretch}{1.}
\caption{
Log of Observations.
\vspace*{3mm}
\label{table_obslog}
}
\begin{tabular}{lccccccc}
\tableline \tableline
\hii & Date & & & \phn\phn $t_{\rm total}$ \phn\phn & & & RMS \\ 
Region & (UT 2003) & Grating & $N_{\rm exp}$ & (s) & 
$\langle X \rangle$ & \phn \othreea \phn & (mag) \\
(1) & (2) & (3) & (4) & (5) & (6) & (7) & (8) \\
\tableline
%
HM \phn2 & 28 Aug & \#11 & 1 $\times$ 1200 & 1200 & 1.24 & no & 0.034 \\
HM \phn2 & 31 Aug & \#7 & 3 $\times$ 1200 & 3600 & 1.21 & no & 0.025 \\
HM \phn7 & 26 Aug & \#11 & 3 $\times$ 1200 & 3600 & 1.06 & yes & 0.030 \\
HM \phn8 & 26 Aug & \#11 & 7 $\times$ 1200 & 8400 & 1.20 & no & 0.030 \\
HM \phn9 & 26 Aug & \#11 & 7 $\times$ 1200 & 8400 & 1.20 & yes & 0.030 \\
HM \phn9 & 31 Aug & \#7 & 3 $\times$ 1200 & 3600 & 1.21 & no & 0.025 \\
HM 12 & 26 Aug & \#11 & 3 $\times$ 1200 & 3600 & 1.06 & no & 0.030 \\
HM 12 & 28 Aug & \#11 & 3 $\times$ 1200 & 3600 & 1.08 & no & 0.034 \\
HM 16 NW & 27 Aug & \#11 & 3 $\times$ 1200 & 3600 & 1.04 & no & 0.029 \\
HM 16 SE & 27 Aug & \#11 & 3 $\times$ 1200 & 3600 & 1.04 & no & 0.029 \\
HM 17 & 28 Aug & \#11 & 3 $\times$ 1200 & 3600 & 1.03 & no & 0.034 \\
HM 18 & 28 Aug & \#11 & 3 $\times$ 1200 & 3600 & 1.03 & no & 0.034 \\
HM 18a & 28 Aug & \#11 & 1 $\times$ 1200 & 1200 & 1.06 & no & 0.034 \\
HM 18b & 28 Aug & \#11 & 1 $\times$ 1200 & 1200 & 1.06 & no & 0.034 \\
HM 19 & 27 Aug & \#11 & 3 $\times$ 1200 & 3600 & 1.04 & no & 0.029 \\
HM 19 & 28 Aug & \#11 & 3 $\times$ 1200 & 3600 & 1.08 & no & 0.034 \\
HM 21 & 28 Aug & \#11 & 3 $\times$ 1200 & 3600 & 1.08 & no & 0.034 \\
\tableline
\vspace*{1mm}
\end{tabular}
%
\tablecomments{
Col.~(1): \hii\ region, following the naming convention by \cite{hm95}.
Col.~(2): Date of observation.
Col.~(3): Grating.
Col.~(4): Number of exposures obtained and the length of each
exposure in seconds. 
Col.~(5): Total exposure time.
Col.~(6): Mean effective airmass.
Col.~(7): \othreea\ detection.
Col.~(8): Relative root--mean--square error in the sensitivity
function obtained from observations of standard stars.
}
\end{center}
\end{table}


\begin{table}
\small 
\tablenum{4a}
\begin{center}
\renewcommand{\arraystretch}{0.95} 
\caption{
Line ratios and properties for \hii\ regions HM~2 and HM~7.
\vspace*{3mm}
\label{table_data1}
}
\begin{tabular}{rccccccc}
\tableline \tableline
& & \multicolumn{2}{c}{HM 2 (gr\#7)} & 
\multicolumn{2}{c}{HM 2 (gr \#11)} &
\multicolumn{2}{c}{HM 7 (gr \#11)} \\
\multicolumn{1}{c}{Wavelength (\AA)} &
\multicolumn{1}{c}{$f(\lambda)$} &
\multicolumn{1}{c}{$F$} & \multicolumn{1}{c}{$I$} &
\multicolumn{1}{c}{$F$} & \multicolumn{1}{c}{$I$} &
\multicolumn{1}{c}{$F$} & \multicolumn{1}{c}{$I$} \\
\tableline
$[\rm{O\;II}]\;3727$ & $+0.325$ &
	$173 \pm 13$ & $198 \pm 29$ &
	$139.0 \pm 8.9$ & $136 \pm 27$ &
	$116.7 \pm 1.9$ & $116.1 \pm 1.9$ 
\\
${\rm H}11\;3772$ & $+0.316$ &
	\nodata & \nodata &
	\nodata & \nodata &
	$3.4 \pm 1.4$ & $4.2 \pm 1.4$
\\
${\rm H}10\;3799$ & $+0.310$ &
	\nodata & \nodata &
	\nodata & \nodata &
	$5.3 \pm 1.6$ & $6.2 \pm 1.5$ 
\\
${\rm H}9\;3835$ & $+0.302$ &
	\nodata & \nodata &
	\nodata & \nodata &
	$7.65 \pm 0.91$ & $8.77 \pm 0.91$
\\
$[{\rm Ne\;III}]\;3869$ & $+0.294$ &
	$55.3 \pm 7.6$ & $60 \pm 13$ &
	$48.8 \pm 5.2$ & $48 \pm 11$ &
	$39.25 \pm 0.95$ & $39.06 \pm 0.95$
\\
${\rm H}8 + {\rm He\;I}\;3889$ & $+0.289$ &
	\nodata & \nodata &
	\nodata & \nodata &
	$22.33 \pm 0.87$ & $23.54 \pm 0.87$
\\
${\rm H}\epsilon + {\rm He\;I}\;3970$\tablenotemark{a} & $+0.269$ &
	$36.1 \pm 5.6$ & $42 \pm 10$ &
	$21.2 \pm 4.4$ & $25.2 \pm 9.5$ &
	$27.84 \pm 0.66$ & $28.92 \pm 0.66$
\\
${\rm He\;I}\;4027$ & $+0.253$ &
	\nodata & \nodata &
	\nodata & \nodata &
	$1.96 \pm 0.48$ & $1.95 \pm 0.48$
\\
$[{\rm S\;II}]\;4068$ & $+0.241$ &
	\nodata & \nodata &
	\nodata & \nodata &
	$1.00 \pm 0.59$ & $1.00 \pm 0.59$
\\
${\rm H}\delta\;4101$ & $+0.232$ &
	$38.1 \pm 6.1$ & $43 \pm 11$ &
	$18.4 \pm 2.5$ & $21.6 \pm 6.2$ &
	$27.49 \pm 0.77$ & $28.43 \pm 0.77$
\\
${\rm H}\gamma\;4340$ & $+0.158$ &
	$42.9 \pm 5.3$ & $46.8 \pm 9.6$ &
	$48.6 \pm 3.7$ & $51 \pm 11$ &
	$47.2 \pm 1.0$ & $47.8 \pm 1.0$
\\
$[{\rm O\;III}]\;4363$ & $+0.151$ &
	\nodata & \nodata &
	$< 12.3$ & $< 12.0$ &
	$7.95 \pm 0.80$ & $7.91 \pm 0.80$
\\
${\rm He\;I}\;4388$ & $+0.143$ &
	\nodata & \nodata &
	\nodata & \nodata &
	$0.63 \pm 0.81$ & $0.63 \pm 0.81$
\\
${\rm He\;I}\;4471$ & $+0.116$ &
	\nodata & \nodata &
	\nodata & \nodata &
	$3.56 \pm 0.25$ & $3.54 \pm 0.25$
\\
$[{\rm Ar\;IV}] + {\rm He\;I}\;4713$ & $+0.042$ &
	\nodata & \nodata &
	\nodata & \nodata &
	$0.98 \pm 0.17$ & $0.98 \pm 0.17$
\\
$[{\rm Ar\;IV}]\;4740$ & $+0.034$ &
	\nodata & \nodata &
	\nodata & \nodata &
	$0.34 \pm 0.14$ & $0.34 \pm 0.14$
\\
${\rm H}\beta\;4861$ & \phs0.000 &
	$100.0 \pm 5.7$ & $100.0 \pm 6.8$ &
	$100 \pm 11$ & $100 \pm 12$ &
	$100.0 \pm 3.7$ & $100.0 \pm 3.6$
\\
$[{\rm O\;III}]\;4959$ & $-0.026$ &
	$187.9 \pm 5.3$ & $183 \pm 20$ &
	$213 \pm 12$ & $209 \pm 39$ &
	$133.9 \pm 9.1$ & $133.2 \pm 9.1$
\\
$[{\rm O\;III}]\;5007$ & $-0.038$ &
	$529.7 \pm 6.4$ & $516 \pm 52$ &
	$568 \pm 15$ & $556 \pm 92$ &
	$390 \pm 12$ & $388 \pm 12$
\\
${\rm He\;I}\;5876$ & $-0.204$ &
	\nodata & \nodata &
	\nodata & \nodata &
	$10.45 \pm 0.38$ & $10.40 \pm 0.38$
\\
$[{\rm O\;I}]\;6300$ & $-0.264$ &
	\nodata & \nodata &
	\nodata & \nodata &
	$2.89 \pm 0.31$\tablenotemark{b} & $2.88 \pm 0.31$
\\
$[{\rm O\;I}]\;6363$ & $-0.272$ &
	\nodata & \nodata &
	\nodata & \nodata &
	$0.52 \pm 0.24$ & $0.51 \pm 0.24$
\\
${\rm H}\alpha\;6563$ & $-0.299$ &
	\nodata & \nodata &
	$280.0 \pm 8.7$ & $276 \pm 47$ &
	$270.5 \pm 7.2$ & $269.4 \pm 7.2$
\\
$[{\rm N\;II}]\;6583$ & $-0.301$ &
	\nodata & \nodata &
	\nodata & \nodata &
	$4.84 \pm 0.59$ & $4.81 \pm 0.59$
\\
${\rm He\;I}\;6678$ & $-0.314$ &
	\nodata & \nodata &
	\nodata & \nodata &
	$3.11 \pm 0.30$ & $3.09 \pm 0.30$
\\
$[{\rm S\;II}]\;6716$ & $-0.319$ &
	\nodata & \nodata &
	\nodata & \nodata &
	$7.57 \pm 0.44$ & $7.53 \pm 0.44$
\\
$[{\rm S\;II}]\;6731$ & $-0.321$ &
	\nodata & \nodata &
	\nodata & \nodata &
	$4.91 \pm 0.38$ & $4.89 \pm 0.38$
\\
${\rm He\;I}\;7065$ & $-0.366$ &
	\nodata & \nodata &
	\nodata & \nodata &
	$2.95 \pm 0.21$ & $2.94 \pm 0.21$
\\
$[{\rm Ar\;III}]\;7136$ & $-0.375$ &
	\nodata & \nodata &
	\nodata & \nodata &
	$5.87 \pm 0.23$ & $5.84 \pm 0.23$
\\
$[{\rm O\;II}]\;7320$ & $-0.400$ &
	\nodata & \nodata &
	\nodata & \nodata &
	$1.51 \pm 0.22$ & $1.50 \pm 0.22$
\\
$[{\rm O\;II}]\;7330$ & $-0.401$ &
	\nodata & \nodata &
	\nodata & \nodata &
	$1.84 \pm 0.23$ & $1.83 \pm 0.23$
\\[1mm]
\multicolumn{2}{c}{$F(\hbeta)$ (ergs s$^{-1}$ cm$^{-2}$)} & 
	\multicolumn{2}{c}{$(4.54 \pm 0.26) \times 10^{-16}$} &
	\multicolumn{2}{c}{$(2.47 \pm 0.27) \times 10^{-15}$} &
	\multicolumn{2}{c}{$(4.63 \pm 0.17) \times 10^{-15}$} 
\\
\multicolumn{2}{c}{EW$_{\rm e}$(\hbeta) (\AA)} &
	\multicolumn{2}{c}{$121 \pm 20$} &
	\multicolumn{2}{c}{$98 \pm 22$} &
	\multicolumn{2}{c}{$403 \pm 114$} 
\\
\multicolumn{2}{c}{Derived $E(B-V)$ (mag)$\,$\tablenotemark{c}} &
	\multicolumn{2}{c}{$+0.087 \pm 0.395$} &
	\multicolumn{2}{c}{$-0.038 \pm 0.165$} &
	\multicolumn{2}{c}{$-0.061 \pm 0.049$} 
\\
\multicolumn{2}{c}{$c(\hbeta)$} &
	\multicolumn{2}{c}{\nodata} &
	\multicolumn{2}{c}{\nodata} &
	\multicolumn{2}{c}{0} 
\\
\multicolumn{2}{c}{Adopted $A_V$ (mag)} &
	\multicolumn{2}{c}{$+0.28$} &
	\multicolumn{2}{c}{0} &
	\multicolumn{2}{c}{0} 
\\
\multicolumn{2}{c}{EW$_{\rm abs}$ (\AA)} &
	\multicolumn{2}{c}{2} &
	\multicolumn{2}{c}{2} &
	\multicolumn{2}{c}{2} 
\\
\tableline
\end{tabular}
\tablenotetext{a}{
Blended with [\ion{Ne}{3}]$\lambda$ 3967.
}
\tablenotetext{b}{
Blended with [\ion{S}{3}]$\lambda$ 6312.
}
\tablenotetext{c}{
HM~2 (gr\#7) : derived from $F$(\hgamma)/$F$(\hbeta);
HM~2 and HM~7 (gr\#11) : derived from $F$(\halpha)/$F$(\hbeta).
}
\tablecomments{
Emission lines are listed in \AA.
$F$ is the observed flux ratio with respect to \hbeta.
$I$ is the corrected intensity ratio, corrected for the adopted
reddening listed, and for underlying Balmer absorption.
The uncertainties in the observed line ratios account for the
uncertainties in the fits to the line profiles, the surrounding
continua, and the relative uncertainty in the sensitivity function
listed in Table~\ref{table_obslog}. 
Flux uncertainties in the \hbeta\ reference line are not included.
Uncertainties in the corrected line ratios account for uncertainties
in the specified line and in the \hbeta\ reference line.
The reddening function, $f(\lambda)$, from
Equation~(\ref{eqn_corr}) is given.
Also listed are the observed \hbeta\ flux; the equivalent width of
\hbeta\ in emission, EW$_{\rm e}$(\hbeta); derived values of the
reddenings from SNAP using Equation~(\ref{eqn_corrthesis}). 
Where \othreea\ is measured, simultaneous solutions for the
logarithmic reddening, $c(\hbeta)$, from Equation~(\ref{eqn_corr}) and
the equivalent width of the underlying Balmer absorption,
EW$_{\rm abs}$ are listed.
The adopted value of the extinction in $V$, $A_V$, is listed.
Where \othreea\ is not measured, the equivalent width of the
underlying Balmer absorption was set to 2~\AA.
}
\end{center}
\end{table}


\begin{table}
\small  
\tablenum{4b}
\begin{center}
\renewcommand{\arraystretch}{1.0} 
\caption{
Line ratios and properties for \hii\ region HM~9.
\vspace*{3mm}
\label{table_data2}
}
\begin{tabular}{rccccccc}
\tableline \tableline
& & \multicolumn{2}{c}{HM 9 ap1 (gr\#7)} & 
\multicolumn{2}{c}{HM 9 ap2\tablenotemark{a} (gr \#7)} &
\multicolumn{2}{c}{HM 9 ap3 (gr \#7)} \\
\multicolumn{1}{c}{Wavelength (\AA)} &
\multicolumn{1}{c}{$f(\lambda)$} &
\multicolumn{1}{c}{$F$} & \multicolumn{1}{c}{$I$} &
\multicolumn{1}{c}{$F$} & \multicolumn{1}{c}{$I$} &
\multicolumn{1}{c}{$F$} & \multicolumn{1}{c}{$I$} \\
\tableline
$[\rm{O\;II}]\;3727$ & $+0.325$ &
	$264.5 \pm 6.2$ & $261 \pm 20$ &
	$301.6 \pm 6.7$ & $296 \pm 22$ &
	$407 \pm 16$ & $397 \pm 47$
\\
$[{\rm Ne\;III}]\;3869$ & $+0.294$ &
	$34.8 \pm 3.7$ & $34.3 \pm 5.1$ &
	$34.0 \pm 4.4$ & $33.4 \pm 5.7$ &
	$46 \pm 11$ & $45 \pm 14$
\\
${\rm H}\epsilon + {\rm He\;I}\;3970$ & $+0.269$ &
	$23.8 \pm 2.5$ & $26.6 \pm 4.3$ &
	$23.1 \pm 2.5$ & $26.4 \pm 4.4$ &
	$44.6 \pm 8.1$& $48 \pm 13$
\\
${\rm H}\delta\;4101$ & $+0.232$ &
	$25.1 \pm 2.2$ & $27.8 \pm 4.0$&
	$24.5 \pm 2.2$ & $27.5 \pm 4.0$ &
	$26.5 \pm 6.5$ & $30 \pm 11$
\\
${\rm H}\gamma\;4340$ & $+0.158$ &
	$45.7 \pm 2.3$ & $47.4 \pm 4.8$ &
	$47.8 \pm 2.5$ & $49.5 \pm 5.0$ &
	$54.7 \pm 6.9$ & $56 \pm 12$
\\
$[{\rm O\;III}]\;4363$ & $+0.151$ &
	$< 7.3$ & $< 7.2$ &
	$< 8.3$ & $< 8.1$ &
	\nodata & \nodata
\\
${\rm H}\beta\;4861$ & \phs0.000 &
	$100.0 \pm 2.2$ & $100.0 \pm 3.8$ &
	$100.0 \pm 2.2$ & $100.0 \pm 3.8$ &
	$100.0 \pm 5.8$ & $100.0 \pm 6.9$
\\
$[{\rm O\;III}]\;4959$ & $-0.026$ &
	$102.6 \pm 2.9$ & $101.1 \pm 8.0$ &
	$93.9 \pm 2.7$ & $92.3 \pm 7.4$ &
	$70.7 \pm 5.3$ & $69 \pm 10$
\\
$[{\rm O\;III}]\;5007$ & $-0.038$ &
	$305.0 \pm 3.5$ & $300 \pm 21$ &
	$279.5 \pm 3.3$ & $275 \pm 19$ &
	$203.1 \pm 6.4$ & $198 \pm 22$
\\[1mm]
\multicolumn{2}{c}{$F(\hbeta)$ (ergs s$^{-1}$ cm$^{-2}$)} & 
	\multicolumn{2}{c}{$(9.70 \pm 0.21) \times 10^{-16}$} &
	\multicolumn{2}{c}{$(1.608 \pm 0.036) \times 10^{-15}$} &
	\multicolumn{2}{c}{$(2.87 \pm 0.17) \times 10^{-16}$} 
\\
\multicolumn{2}{c}{EW$_{\rm e}$(\hbeta) (\AA)} &
	\multicolumn{2}{c}{$132.5 \pm 9.4$} &
	\multicolumn{2}{c}{$114.3 \pm 7.2$} &
	\multicolumn{2}{c}{$80 \pm 10$} 
\\
\multicolumn{2}{c}{Derived $E(B-V)$ (mag)\tablenotemark{b}} &
	\multicolumn{2}{c}{$-0.025 \pm 0.195$} &
	\multicolumn{2}{c}{$-0.108 \pm 0.197$} &
	\multicolumn{2}{c}{$-0.338 \pm 0.409$} 
\\
\multicolumn{2}{c}{$c(\hbeta)$} &
	  \multicolumn{2}{c}{\nodata} &
	  \multicolumn{2}{c}{\nodata} &
	  \multicolumn{2}{c}{\nodata} 
\\
\multicolumn{2}{c}{Adopted $A_V$ (mag)} &
	\multicolumn{2}{c}{0} &
	\multicolumn{2}{c}{0} &
	\multicolumn{2}{c}{0} 
\\
\multicolumn{2}{c}{EW$_{\rm abs}$ (\AA)} &
	\multicolumn{2}{c}{2} &
	\multicolumn{2}{c}{2} &
	\multicolumn{2}{c}{2} 
\\[1mm]
\tableline
& & \multicolumn{2}{c}{HM 9 ap1 (gr\#11)} & 
\multicolumn{2}{c}{HM 9 ap2\tablenotemark{a} (gr \#11)} &
\multicolumn{2}{c}{HM 9 ap3 (gr \#11)} \\
\multicolumn{1}{c}{Wavelength (\AA)} &
\multicolumn{1}{c}{$f(\lambda)$} &
\multicolumn{1}{c}{$F$} & \multicolumn{1}{c}{$I$} &
\multicolumn{1}{c}{$F$} & \multicolumn{1}{c}{$I$} &
\multicolumn{1}{c}{$F$} & \multicolumn{1}{c}{$I$} \\
\tableline
$[\rm{O\;II}]\;3727$ & $+0.325$ &
	$208.7 \pm 4.2$ & $202 \pm 11$ &
	$215.9 \pm 4.2$ & $209 \pm 12$ &
	$213.4 \pm 4.8$ & $204 \pm 12$ 
\\
$[{\rm Ne\;III}]\;3869$ & $+0.294$ &
	$40.4 \pm 1.7$ & $39.1 \pm 2.5$ &
	$42.7 \pm 1.2$ & $41.3 \pm 2.3$ &
	$48.0 \pm 2.3$ & $45.9 \pm 3.1$ 
\\
${\rm H}8 + {\rm He\;I}\;3889$ & $+0.289$ &
	$15.7 \pm 1.4$ & $23.0 \pm 1.7$ &
	$16.5 \pm 1.0$ & $24.3 \pm 1.5$ &
	$13.2 \pm 1.8$ & $24.5 \pm 2.1$
\\
${\rm H}\epsilon + {\rm He\;I}\;3970$ & $+0.269$ &
	$22.7 \pm 1.5$ & $29.1 \pm 1.9$ &
	$20.1 \pm 1.1$ & $27.2 \pm 1.6$ &
	$17.0 \pm 2.0$ & $27.4 \pm 2.3$
\\
${\rm H}\delta\;4101$ & $+0.232$ &
	$20.4 \pm 1.5$ & $26.3 \pm 1.8$ &
	$19.9 \pm 1.1$ & $26.1 \pm 1.4$ &
	$16.8 \pm 1.6$ & $25.8 \pm 1.8$
\\
${\rm H}\gamma\;4340$ & $+0.158$ &
	$45.6 \pm 1.5$ & $49.1 \pm 1.9$ &
	$45.3 \pm 1.4$ & $48.9 \pm 1.8$ &
	$43.5 \pm 2.2$ & $48.6 \pm 2.5$
\\
$[{\rm O\;III}]\;4363$ & $+0.151$ &
	$7.1 \pm 1.1$ & $6.9 \pm 1.1$ &
	$6.5 \pm 1.1$ & $6.3 \pm 1.1$ &
	$5.7 \pm 1.7$ & $5.4 \pm 1.6$
\\
${\rm He\;I}\;4471$ & $+0.116$ &
	\nodata & \nodata &
	$3.36 \pm 0.67$ & $3.25 \pm 0.65$ &
	$3.67 \pm 0.90$ & $3.50 \pm 0.86$
\\
${\rm H}\beta\;4861$ & \phs0.000 &
	$100.0 \pm 3.9$ & $100.0 \pm 3.8$ &
	$100.0 \pm 3.8$ & $100.0 \pm 3.7$ &
	$100.0 \pm 4.2$ & $100.0 \pm 4.0$
\\
$[{\rm O\;III}]\;4959$ & $-0.026$ &
	$143 \pm 11$ & $138 \pm 11$ &
	$140 \pm 11$ & $135 \pm 11$ &
	$135.1 \pm 9.7$ & $128.9 \pm 9.3$
\\
$[{\rm O\;III}]\;5007$ & $-0.038$ &
	$420 \pm 14$ & $406 \pm 14$ &
	$407 \pm 14$ & $394 \pm 14$ &
	$392 \pm 12$ & $374 \pm 12$
\\
${\rm He\;I}\;5876$ & $-0.204$ &
	$10.6 \pm 1.1$ & $10.3 \pm 1.1$ &
	$10.39 \pm 0.78$ & $10.05 \pm 0.82$ &
	$10.9 \pm 1.5$ & $10.4 \pm 1.5$
\\
${\rm H}\alpha\;6563$ & $-0.299$ &
	$279.9 \pm 8.7$ & $272 \pm 16$ &
	$287.3 \pm 9.4$ & $279 \pm 16$ &
	$295.0 \pm 9.2$ & $282 \pm 16$
\\
$[{\rm N\;II}]\;6583$ & $-0.301$ &
	$< 4.6$ & $< 4.5$ &
	$7.7 \pm 2.3$ & $7.5 \pm 2.3$ &
	$8.7 \pm 1.8$ & $8.3 \pm 1.8$
\\
$[{\rm S\;II}]\;6716,6731$ & $-0.320$ &
	$29.8 \pm 1.8$ & $28.8 \pm 2.3$ &
	$35.2 \pm 1.7$ & $34.0 \pm 2.4$ &
	$41.9 \pm 2.6$ & $39.9 \pm 3.2$
\\
$[{\rm Ar\;III}]\;7136$ & $-0.375$ &
	$8.7 \pm 1.1$ & $8.6 \pm 1.6$ &
	$8.00 \pm 0.89$ & $7.9 \pm 1.3$ &
	$8.9 \pm 2.4$ & $8.7 \pm 2.8$
\\[1mm]
\multicolumn{2}{c}{$F(\hbeta)$ (ergs s$^{-1}$ cm$^{-2}$)} & 
	\multicolumn{2}{c}{$(2.67 \pm 0.11) \times 10^{-16}$} &
	\multicolumn{2}{c}{$(6.52 \pm 0.25) \times 10^{-16}$} &
	\multicolumn{2}{c}{$(1.942 \pm 0.082) \times 10^{-16}$}
\\
\multicolumn{2}{c}{EW$_{\rm e}$(\hbeta) (\AA)} &
	\multicolumn{2}{c}{$123 \pm 11$} &
	\multicolumn{2}{c}{$111.0 \pm 8.7$} &
	\multicolumn{2}{c}{$84.4 \pm 6.1$}
\\
\multicolumn{2}{c}{Derived $E(B-V)$ (mag)\tablenotemark{c}} &
	\multicolumn{2}{c}{$-0.036 \pm 0.096$} &
	\multicolumn{2}{c}{$-0.012 \pm 0.096$} &
	\multicolumn{2}{c}{$+0.009 \pm 0.099$}
\\
\multicolumn{2}{c}{$c(\hbeta)$} &
	\multicolumn{2}{c}{$-0.04 \pm 0.07$} &
	\multicolumn{2}{c}{$-0.01 \pm 0.07$} &
	\multicolumn{2}{c}{$+0.00 \pm 0.07$}
\\
\multicolumn{2}{c}{Adopted $A_V$ (mag)} &
	\multicolumn{2}{c}{0} &
	\multicolumn{2}{c}{0} &
	\multicolumn{2}{c}{0}
\\
\multicolumn{2}{c}{EW$_{\rm abs}$ (\AA)} &
	\multicolumn{2}{c}{$4.1 \pm 2.0$} &
	\multicolumn{2}{c}{$3.8 \pm 1.4$} &
	\multicolumn{2}{c}{$4.1 \pm 1.3$}
\\
\tableline
\end{tabular}
\vspace*{1mm}
\tablenotetext{a}{
For the given grating setting, the definition for aperture ``2''
encompasses regions 1 and 3. 
}
\tablenotetext{b}{
Derived from $F$(\hgamma)/$F$(\hbeta) ratios.
}
\tablenotetext{c}{
Derived from $F$(\halpha)/$F$(\hbeta) ratios.
}
\tablecomments{
Grating \#7 : long-slit aligned with HM~2.
Grating \#11 : long-slit aligned with HM~8.
Thus, spectra for the first aperture in both grating settings are
likely at the same position within the HM~9 nebula.
See also Table~\ref{table_data1} for comments.
}
\end{center}
\end{table}

\begin{table}
\small 
\tablenum{4c}
\begin{center}
\renewcommand{\arraystretch}{1.} 
\caption{
Line ratios and properties for \hii\ regions HM~8, 12, 16, and 17.
\vspace*{3mm}
\label{table_data3}
}
\begin{tabular}{rccccccc}
\tableline \tableline
& & \multicolumn{2}{c}{HM 8 (gr\#11)} & 
\multicolumn{2}{c}{HM 12 (gr \#11)\tablenotemark{a}} & 
\multicolumn{2}{c}{HM 12 (gr \#11)\tablenotemark{b}} \\ 
\multicolumn{1}{c}{Wavelength (\AA)} &
\multicolumn{1}{c}{$f(\lambda)$} &
\multicolumn{1}{c}{$F$} & \multicolumn{1}{c}{$I$} &
\multicolumn{1}{c}{$F$} & \multicolumn{1}{c}{$I$} &
\multicolumn{1}{c}{$F$} & \multicolumn{1}{c}{$I$} \\
\tableline
$[\rm{O\;II}]\;3727$ & $+0.325$ &
	$350 \pm 15$ & $339 \pm 43$ &
	$445 \pm 12$ & $432 \pm 42$ &
	$289.9 \pm 9.9$ & $281 \pm 34$
\\
$[{\rm Ne\;III}]\;3869$ & $+0.294$ &
	\nodata & \nodata &
	\nodata & \nodata &
	$37.1 \pm 5.7$ & $36.0 \pm 8.3$ 
\\
${\rm H}\delta\;4101$ & $+0.232$ &
	\nodata & \nodata &
	\nodata & \nodata &
	$22.2 \pm 6.0$ & $25 \pm 10$
\\
${\rm H}\gamma\;4340$ & $+0.158$ &
	$39.7 \pm 4.9$ & $40.7 \pm 8.6$ &
	$33.6 \pm 3.3$ & $36.9 \pm 6.3$ &
	$49.2 \pm 5.6$ & $51 \pm 10$
\\
${\rm H}\beta\;4861$ & \phs0.000 &
	$100.0 \pm 6.5$ & $100.0 \pm 7.5$ &
	$100.0 \pm 4.5$ & $100.0 \pm 5.6$ &
	$100.0 \pm 6.1$ & $100.0 \pm 7.3$ 
\\
$[{\rm O\;III}]\;4959$ & $-0.026$ &
	$46.9 \pm 5.6$ & $45.4 \pm 9.0$ &
	$40.1 \pm 4.0$ & $38.9 \pm 6.2$ &
	$75.6 \pm 6.0$ & $73 \pm 12$
\\
$[{\rm O\;III}]\;5007$ & $-0.038$ &
	$144.1 \pm 7.0$ & $140 \pm 18$ &
	$114.8 \pm 4.8$ & $111 \pm 12$ &
	$257.0 \pm 8.0$ & $249 \pm 30$
\\
${\rm He\;I}\;5876$ & $-0.204$ &
	\nodata & \nodata &
	\nodata & \nodata &
	$15.0 \pm 3.3$ & $14.6 \pm 4.4$
\\
${\rm H}\alpha\;6563$ & $-0.299$ &
	$266 \pm 10$ & $264 \pm 33$ &
	$284 \pm 10$ & $277 \pm 29$ &
	$288.5 \pm 9.1$ & $282 \pm 34$
\\
$[{\rm N\;II}]\;6583$ & $-0.301$ &
	\nodata & \nodata &
	$< 7.7$ & $< 7.5$ &
	\nodata & \nodata
\\
$[{\rm S\;II}]\;6716,6731$ & $-0.320$ &
	$29.6 \pm 5.1$ & $28.7 \pm 7.2$ &
	$69.2 \pm 7.2$ & $67 \pm 11$ &
	$37.0 \pm 4.6$ & $36.0 \pm 7.3$
\\[1mm]
\multicolumn{2}{c}{$F(\hbeta)$ (ergs s$^{-1}$ cm$^{-2}$)} & 
	\multicolumn{2}{c}{$(4.50 \pm 0.29) \times 10^{-17}$} &
	\multicolumn{2}{c}{$(1.85 \pm 0.08) \times 10^{-16}$} &
	\multicolumn{2}{c}{$(3.43 \pm 0.21) \times 10^{-16}$}
\\
\multicolumn{2}{c}{EW$_{\rm e}$(\hbeta) (\AA)} &
	\multicolumn{2}{c}{$62.2 \pm 5.8$} &
	\multicolumn{2}{c}{$64.9 \pm 4.3$} &
	\multicolumn{2}{c}{$66.5 \pm 6.2$}
\\
\multicolumn{2}{c}{Derived $E(B-V)$ (mag)} &
	\multicolumn{2}{c}{$-0.083 \pm 0.124$} &
	\multicolumn{2}{c}{$-0.035 \pm 0.106$} &
	\multicolumn{2}{c}{$-0.016 \pm 0.121$} 
\\
\multicolumn{2}{c}{$c(\hbeta)$} &
	  \multicolumn{2}{c}{\nodata} &
	  \multicolumn{2}{c}{\nodata} &
	  \multicolumn{2}{c}{\nodata} 
\\
\multicolumn{2}{c}{Adopted $A_V$ (mag)} &
	\multicolumn{2}{c}{0} &
	\multicolumn{2}{c}{0} &
	\multicolumn{2}{c}{0} 
\\
\multicolumn{2}{c}{EW$_{\rm abs}$ (\AA)} &
	\multicolumn{2}{c}{2} &
	\multicolumn{2}{c}{2} &
	\multicolumn{2}{c}{2} 
\\[1mm]
\tableline
& & \multicolumn{2}{c}{HM 16 NW (gr\#11)} & 
\multicolumn{2}{c}{HM 16 SE (gr \#11)} &
\multicolumn{2}{c}{HM 17 (gr \#11)} \\
\multicolumn{1}{c}{Wavelength (\AA)} &
\multicolumn{1}{c}{$f(\lambda)$} &
\multicolumn{1}{c}{$F$} & \multicolumn{1}{c}{$I$} &
\multicolumn{1}{c}{$F$} & \multicolumn{1}{c}{$I$} &
\multicolumn{1}{c}{$F$} & \multicolumn{1}{c}{$I$} \\
\tableline
$[\rm{O\;II}]\;3727$ & $+0.325$ &
	$309.0 \pm 6.0$ & $431 \pm 33$ &
	$470 \pm 19$ & $700 \pm 94$ &
	$448 \pm 16$ & $428 \pm 55$
\\
${\rm H}\epsilon + {\rm He\;I}\;3970$ & $+0.269$ &
	$13.2 \pm 1.5$ & $19.9 \pm 3.6$ &
	\nodata & \nodata &
	\nodata & \nodata
\\
${\rm H}\delta\;4101$ & $+0.232$ &
	$17.1 \pm 1.5$ & $23.4 \pm 3.4$ &
	\nodata & \nodata &
	\nodata & \nodata
\\
${\rm H}\gamma\;4340$ & $+0.158$ &
	$41.8 \pm 1.6$ & $48.9 \pm 4.5$ &
	$46.5 \pm 7.6$ & $60 \pm 17$ &
	$35.7 \pm 6.1$ & $39 \pm 11$
\\
$[{\rm O\;III}]\;4363$ & $+0.151$ &
	$< 4.8$ & $< 5.3$ &
	\nodata & \nodata &
	\nodata & \nodata
\\
${\rm H}\beta\;4861$ & \phs0.000 &
	$100.0 \pm 2.8$ & $100.0 \pm 4.1$ &
	$100.0 \pm 7.2$ & $100.0 \pm 8.3$ &
	$100.0 \pm 6.7$ & $100.0 \pm 7.9$
\\
$[{\rm O\;III}]\;4959$ & $-0.026$ &
	$71.1 \pm 4.6$ & $68.8 \pm 7.7$ &
	$10.8 \pm 5.5$ & $9.9 \pm 5.9$ &
	\nodata & \nodata
\\
$[{\rm O\;III}]\;5007$ & $-0.038$ &
	$214.2 \pm 6.0$ & $205 \pm 17$ &
	$13.2 \pm 5.6$ & $12.0 \pm 6.2$ &
	$56.6 \pm 5.2$ & $54.0 \pm 9.6$
\\
${\rm He\;I}\;5876$ & $-0.204$ &
	$10.6 \pm 1.6$ & $8.8 \pm 1.7$ &
	\nodata & \nodata &
	\nodata & \nodata
\\
${\rm H}\alpha\;6563$ & $-0.299$ &
	$370.5 \pm 8.3$ & $286 \pm 23$ &
	$416 \pm 13$ & $286 \pm 36$ &
	$256.6 \pm 9.8$ & $248 \pm 33$
\\
$[{\rm N\;II}]\;6583$ & $-0.301$ &
	$14.2 \pm 3.2$ & $10.9 \pm 3.0$ &
	\nodata & \nodata &
	\nodata & \nodata
\\
$[{\rm S\;II}]\;6716,6731$ & $-0.320$ &
	$26.5 \pm 3.2$ & $20.1 \pm 3.3$ &
	$64.8 \pm 9.2$ & $43.3 \pm 9.9$ &
	$42.6 \pm 5.7$ & $40.6 \pm 8.9$
\\[1mm]
\multicolumn{2}{c}{$F(\hbeta)$ (ergs s$^{-1}$ cm$^{-2}$)} & 
	\multicolumn{2}{c}{$(6.86 \pm 0.19) \times 10^{-16}$} &
	\multicolumn{2}{c}{$(1.38 \pm 0.10) \times 10^{-16}$} &
	\multicolumn{2}{c}{$(2.39 \pm 0.16) \times 10^{-16}$}
\\
\multicolumn{2}{c}{EW$_{\rm e}$(\hbeta) (\AA)} &
	\multicolumn{2}{c}{$156 \pm 13$} &
	\multicolumn{2}{c}{$34.6 \pm 2.9$} &
	\multicolumn{2}{c}{$41.7 \pm 3.3$}
\\
\multicolumn{2}{c}{Derived $E(B-V)$ (mag)} &
	\multicolumn{2}{c}{$+0.251 \pm 0.080$} &
	\multicolumn{2}{c}{$+0.332 \pm 0.128$} &
	\multicolumn{2}{c}{$-0.145 \pm 0.133$}
\\
\multicolumn{2}{c}{$c(\hbeta)$} &
	  \multicolumn{2}{c}{\nodata} &
	  \multicolumn{2}{c}{\nodata} &
	  \multicolumn{2}{c}{\nodata}
\\
\multicolumn{2}{c}{Adopted $A_V$ (mag)} &
	\multicolumn{2}{c}{$+0.77$} &
	\multicolumn{2}{c}{$+1.0$} &
	\multicolumn{2}{c}{0}
\\
\multicolumn{2}{c}{EW$_{\rm abs}$ (\AA)} &
	\multicolumn{2}{c}{2} &
	\multicolumn{2}{c}{2} &
	\multicolumn{2}{c}{2}
\\
\tableline
\end{tabular}
\vspace*{1mm}
\tablenotetext{a}{
Observed August 26 (UT); long-slit aligned with HM~7.
}
\tablenotetext{b}{
Observed August 28 (UT); long-slit aligned with HM~19 and HM~21.
}
\tablecomments{
See Table~\ref{table_data1} for comments.
}
\end{center}
\end{table}

\begin{table}
\small    
\tablenum{4d}
\begin{center}
\renewcommand{\arraystretch}{1.0} 
\caption{
Line ratios and properties for \hii\ regions HM~18, 18a, 18b, 19,
and 21.
\vspace*{3mm}
\label{table_data4}
}
\begin{tabular}{rccccccc}
\tableline \tableline
& & \multicolumn{2}{c}{HM 18 (gr\#11)} & 
\multicolumn{2}{c}{HM 18a (gr \#11)} &
\multicolumn{2}{c}{HM 18b (gr \#11)} \\
\multicolumn{1}{c}{Wavelength (\AA)} &
\multicolumn{1}{c}{$f(\lambda)$} &
\multicolumn{1}{c}{$F$} & \multicolumn{1}{c}{$I$} &
\multicolumn{2}{c}{$F$} & 
\multicolumn{2}{c}{$F$} \\
\tableline
$[\rm{O\;II}]\;3727$ & $+0.325$ &
	$790 \pm 41$ & $1000 \pm 200$ &
	\multicolumn{2}{c}{\nodata} &
	\multicolumn{2}{c}{\nodata}
\\
${\rm H}\beta\;4861$ & \phs0.000 &
	$100 \pm 10$ & $100 \pm 14$ &
	\multicolumn{2}{c}{$100 \pm 19$} &
	\multicolumn{2}{c}{$100 \pm 17$}
\\
${\rm H}\alpha\;6563$ & $-0.299$ &
	$618 \pm 18$ & $286 \pm 54$ &
	\multicolumn{2}{c}{$554 \pm 37$} & 
	\multicolumn{2}{c}{$881 \pm 50$} 
\\
$[{\rm S\;II}]\;6716,6731$ & $-0.320$ &
	$92 \pm 14$ & $40 \pm 12$ &
	\multicolumn{2}{c}{\nodata} &
	\multicolumn{2}{c}{\nodata}
\\[1mm]
\multicolumn{2}{c}{$F(\hbeta)$ (ergs s$^{-1}$ cm$^{-2}$)} & 
	\multicolumn{2}{c}{$(9.50 \pm 0.99) \times 10^{-17}$} &
	\multicolumn{2}{c}{$(1.47 \pm .27) \times 10^{-16}$} &
	\multicolumn{2}{c}{$(7.6 \pm 1.3) \times 10^{-17}$}
\\
\multicolumn{2}{c}{EW$_{\rm e}$(\hbeta) (\AA)} &
	\multicolumn{2}{c}{$4.61 \pm 0.48$} &
	\multicolumn{2}{c}{$8.8 \pm 1.6$} &
	\multicolumn{2}{c}{$10.7 \pm 1.9$}
\\
\multicolumn{2}{c}{Derived $E(B-V)$ (mag)} &
	\multicolumn{2}{c}{$+0.445 \pm 0.190$} &
	\multicolumn{2}{c}{$+0.481 \pm 0.308$} &
	\multicolumn{2}{c}{$+0.982 \pm 0.277$} 
\\
\multicolumn{2}{c}{$c(\hbeta)$} &
	\multicolumn{2}{c}{\nodata} &
	\multicolumn{2}{c}{\nodata} &
	\multicolumn{2}{c}{\nodata}
\\
\multicolumn{2}{c}{Adopted $A_V$ (mag)} &
	\multicolumn{2}{c}{$+1.37$} &
	\multicolumn{2}{c}{$+1.48$} &
	\multicolumn{2}{c}{$+3.01$} 
\\
\multicolumn{2}{c}{EW$_{\rm abs}$ (\AA)} &
	\multicolumn{2}{c}{2} &
	\multicolumn{2}{c}{2} &
	\multicolumn{2}{c}{2} 
\\[1mm]
\tableline
& & \multicolumn{2}{c}{HM 19 (gr\#11)\tablenotemark{a}} & 
\multicolumn{2}{c}{HM 19 (gr \#11)\tablenotemark{b}} & 
\multicolumn{2}{c}{HM 21 (gr \#11)} \\
\multicolumn{1}{c}{Wavelength (\AA)} &
\multicolumn{1}{c}{$f(\lambda)$} &
\multicolumn{1}{c}{$F$} & \multicolumn{1}{c}{$I$} &
\multicolumn{1}{c}{$F$} & \multicolumn{1}{c}{$I$} &
\multicolumn{1}{c}{$F$} & \multicolumn{1}{c}{$I$} \\
\tableline
$[\rm{O\;II}]\;3727$ & $+0.325$ &
	$350.2 \pm 9.0$ & $467 \pm 48$ &
	$131.6 \pm 4.6$ & $184 \pm 16$ &
	$387.0 \pm 7.9$ & $404 \pm 38$
\\
$[{\rm Ne\;III}]\;3869$ & $+0.294$ &
	\nodata & \nodata &
	$17.1 \pm 3.0$ & $22.4 \pm 4.9$ &
	\nodata & \nodata
\\
${\rm H}\epsilon + {\rm He\;I}\;3970$ & $+0.269$ &
	\nodata & \nodata &
	\nodata & \nodata &
	$12.9 \pm 2.4$ & $18.5 \pm 5.7$
\\
${\rm H}\delta\;4101$ & $+0.232$ &
	\nodata & \nodata &
	$10.6 \pm 1.6$ & $24.0 \pm 6.6$ &
	$20.9 \pm 2.2$ & $26.1 \pm 4.8$
\\
${\rm H}\gamma\;4340$ & $+0.158$ &
	$39.2 \pm 3.3$ & $46.7 \pm 7.5$ &
	$30.0 \pm 1.5$ & $41.7 \pm 4.7$ &
	$43.6 \pm 2.0$ & $47.3 \pm 5.5$
\\
${\rm H}\beta\;4861$ & \phs0.000 &
	$100.0 \pm 5.4$ & $100.0 \pm 6.3$ &
	$100.0 \pm 1.9$ & $100.0 \pm 4.0$ &
	$100.0 \pm 4.0$ & $100.0 \pm 5.4$
\\
$[{\rm O\;III}]\;4959$ & $-0.026$ &
	$67.6 \pm 5.0$ & $65.0 \pm 9.3$ &
	$10.7 \pm 1.4$ & $9.9 \pm 1.8$ &
	$56.9 \pm 3.8$ & $55.2 \pm 7.1$
\\
$[{\rm O\;III}]\;5007$ & $-0.038$ &
	$177.7 \pm 6.1$ & $170 \pm 18$ &
	$36.4 \pm 1.6$ & $33.3 \pm 3.2$ &
	$175.7 \pm 4.9$ & $170 \pm 17$
\\
${\rm H}\alpha\;6563$ & $-0.299$ &
	$363 \pm 14$ & $286 \pm 32$ &
	$397.1 \pm 4.0$ & $286 \pm 22$ &
	$308.1 \pm 7.6$ & $286 \pm 28$
\\
$[{\rm N\;II}]\;6583$ & $-0.301$ &
	\nodata & \nodata &
	$6.7 \pm 3.3$ & $4.8 \pm 2.6$ &
	$< 13$ & $< 12$
\\
$[{\rm S\;II}]\;6716,6731$ & $-0.320$ &
	\nodata & \nodata &
	$8.4 \pm 1.3$ & $5.9 \pm 1.1$ &
	$35.3 \pm 3.9$ & $32.6 \pm 5.6$
\\[1mm]
\multicolumn{2}{c}{$F(\hbeta)$ (ergs s$^{-1}$ cm$^{-2}$)} & 
	\multicolumn{2}{c}{$(3.78 \pm 0.20) \times 10^{-16}$} &
	\multicolumn{2}{c}{$(9.24 \pm 0.17) \times 10^{-16}$} &
	\multicolumn{2}{c}{$(6.70 \pm 0.27) \times 10^{-16}$}
\\
\multicolumn{2}{c}{EW$_{\rm e}$(\hbeta) (\AA)} &
	\multicolumn{2}{c}{$100 \pm 11$} &
	\multicolumn{2}{c}{$35.54 \pm 0.75$} &
	\multicolumn{2}{c}{$71.8 \pm 4.6$}
\\
\multicolumn{2}{c}{Derived $E(B-V)$ (mag)} &
	\multicolumn{2}{c}{$+0.223 \pm 0.113$} &
	\multicolumn{2}{c}{$+0.285 \pm 0.077$} &
	\multicolumn{2}{c}{$+0.050 \pm 0.098$}
\\
\multicolumn{2}{c}{$c(\hbeta)$} &
	\multicolumn{2}{c}{\nodata} &
	\multicolumn{2}{c}{\nodata} &
	\multicolumn{2}{c}{\nodata}
\\
\multicolumn{2}{c}{Adopted $A_V$ (mag)} &
	\multicolumn{2}{c}{$+0.68$} &
	\multicolumn{2}{c}{$+0.87$} &
	\multicolumn{2}{c}{$+0.15$}
\\
\multicolumn{2}{c}{EW$_{\rm abs}$ (\AA)} &
	\multicolumn{2}{c}{2} &
	\multicolumn{2}{c}{2} &
	\multicolumn{2}{c}{2}
\\
\tableline
\end{tabular}
\tablenotetext{a}{
Observed August 27 (UT); long-slit aligned with HM~16 NW
and HM~16 SE.
}
\tablenotetext{b}{
Observed August 28 (UT); long-slit aligned with HM~21.
}
\tablecomments{
See Table~\ref{table_data1} for comments.
}
\end{center}
\end{table}


\begin{table}
\small 
\tablenum{5a}
\begin{center}
\renewcommand{\arraystretch}{1.1}
\caption{
Ionic and total abundances.
\vspace*{3mm}
\label{table_abund1}
}
\begin{tabular}{lccccccc}
\tableline \tableline
Property & HM 2 & HM 2 & HM 7 & HM 8 & HM 9 ap1 & HM 9 ap2 & HM 9 ap3 \\
& 
(gr \#7) & (gr \#11) & (gr \#11) & (gr \#11) & (gr \#7) & (gr \#7) & 
(gr \#7) \\
\tableline
$T_e$(O$^{+2}$) (K) &
\nodata & \phn $< 15800$ \phn & $15350 \pm 760$ & \nodata &
\phn\phn $< 16500$ \phn\phn & \phn\phn $< 18500$ \phn\phn & \nodata
\\
$T_e$(O$^+$) (K) &
\nodata & $< 14100$ & $13750 \pm 680$ & \nodata &
$< 14600$ & $< 16000$ & \nodata
\\
O$^+$/H $(\times 10^5)$ & 
\nodata & $> 1.4$ & $1.30 \pm 0.21$ & \nodata &
$> 2.4$ & $> 2.0$ & \nodata
\\
O$^{+2}$/H $(\times 10^5)$ &
\nodata & $> 5.3$ & $3.92 \pm 0.45$ & \nodata & 
$> 2.5$ & $> 1.7$ & \nodata
\\
O/H $(\times 10^5)$ & 
\nodata & $> 6.7$ & $5.22 \pm 0.50$ & \nodata & 
$> 4.9$ & $> 3.8$ & \nodata
\\
12$+$log(O/H) &
\nodata & $> 7.83$ & 
\phn\phn $7.72 \pm 0.04 \, (^{+0.05}_{-0.06})$ \phn\phn & 
\nodata & $> 7.69$ & $> 7.58$ & \nodata
\\
12$+$log(O/H) M91\tablenotemark{a} &
8.17 & 8.11 & 7.84 & 8.01 &
8.03 & 8.06 & 8.15
\\
12$+$log(O/H) P00\tablenotemark{b} &
7.92 & 7.85 & 7.67 & 8.18 &
7.92 & 7.99 & 8.23
\\
\tableline
Ar$^{+2}$/H $(\times 10^7)$ &
\nodata & \nodata & $2.47 \pm 0.56$ &
\nodata & \nodata & \nodata & \nodata
\\
Ar$^{+3}$/H $(\times 10^7)$ &
\nodata & \nodata & $0.25 \pm 0.15$ &
\nodata & \nodata & \nodata & \nodata
\\
ICF(Ar) &
\nodata & \nodata & 1.06 &
\nodata & \nodata & \nodata & \nodata
\\
Ar/H $(\times 10^7)$ &
\nodata & \nodata & $2.88 \pm 0.61$ &
\nodata & \nodata & \nodata & \nodata
\\
log(Ar/O) &
\nodata & \nodata & $-2.25 \pm 0.10$ &
\nodata & \nodata & \nodata & \nodata
\\
N$^+$/O$^+$ &
\nodata & \nodata & $0.034 \pm 0.004$ &
\nodata & \nodata & \nodata & \nodata
\\
log(N/O) &
\nodata & \nodata & $-1.46 \pm 0.05$ &
\nodata & \nodata & \nodata & \nodata
\\
Ne$^{+2}$/O$^{+2}$ &
\nodata & \nodata & $0.255 \pm 0.027$ &
\nodata & \nodata & \nodata & \nodata
\\
log(Ne/O) &
\nodata & \nodata & $-0.594 \pm 0.046$ &
\nodata & \nodata & \nodata & \nodata
\\
\tableline
\vspace*{1mm}
\end{tabular}
\tablenotetext{a}{
\cite{mcgaugh91} bright-line calibration.
}
\tablenotetext{b}{
\cite{pilyugin00} bright-line calibration.
}
\tablecomments{
Direct oxygen abundances are shown with two uncertainties.
The first uncertainty is the formal uncertainty in the derivation.
In parentheses is the range of possible values, expressed by the
maximum and minimum values of the oxygen abundance.
}
\end{center}
\end{table}


\begin{table}
\small 
\tablenum{5b}
\begin{center}
\renewcommand{\arraystretch}{1.1}
\caption{
Ionic and total abundances (continued).
\vspace*{3mm}
\label{table_abund2}
}
\begin{tabular}{lcccccc}
\tableline \tableline
Property & HM 9 ap1 & HM 9 ap2 & HM 9 ap3 & HM 12\tablenotemark{a} & 
HM 12\tablenotemark{a} & HM 16 NW \\
& (gr \#11) & (gr \#11) & (gr \#11) & (gr \#11) & (gr \#11) & (gr \#11) \\
\tableline
$T_e$(O$^{+2}$) (K) & 
$14100 \pm 1000$ & $13800 \pm 1000$ & $13300 \pm 1800$ & 
\nodata & \nodata & $< 17300$ \\
$T_e$(O$^+$) (K) &
$12890 \pm 910$ & $12650 \pm 950$ & $12300 \pm 1600$ &
\nodata & \nodata & $< 15100$  \\
O$^+$/H ($\times 10^5$) &
$2.80 \pm 0.68$ & $3.10 \pm 0.82$ & $3.4 \pm 1.6$ &
\nodata & \nodata & $> 3.5$ \\
O$^{+2}$/H ($\times 10^5$) &
$5.05 \pm 0.87$ & $5.24 \pm 0.98$ & $5.5 \pm 1.9$ &
\nodata & \nodata & $> 1.5$ \\
O/H ($\times 10^5$) &
$7.9 \pm 1.1$ & $8.3 \pm 1.3$ & $8.9 \pm 2.4$ &
\nodata & \nodata & $> 5.0$ \\
12$+$log(O/H) &
\phn\phn $7.90 \pm 0.06 \, (^{+0.07}_{-0.09})$ \phn & 
\phn $7.92 \pm 0.06 \, (^{+0.08}_{-0.09})$ \phn & 
\phn $7.95 \pm 0.11 \, (^{+0.12}_{-0.18})$ \phn\phn & 
\nodata & \nodata & $> 7.70$ \\
12$+$log(O/H) M91\tablenotemark{b} & 
8.04 & 8.04 & 8.01 &
8.16 & 8.00 & 8.21
\\
12$+$log(O/H) P00\tablenotemark{c} &
7.85 & 7.86 & 7.84 &
8.48 & 7.96 & 8.29 
\\
\tableline
Ar$^{+2}$/H $(\times 10^7)$ &
$4.2 \pm 1.6$ & $4.0 \pm 1.4$ & $4.7 \pm 3.1$ &
\nodata & \nodata & \nodata 
\\
ICF(Ar) &
1.50 & 1.48 & 1.48 &
\nodata & \nodata & \nodata 
\\
Ar/H $(\times 10^7)$ &
$6.3 \pm 2.4$ & $5.9 \pm 2.1$ & $6.9 \pm 4.5$ &
\nodata & \nodata & \nodata 
\\
log(Ar/O) &
$-2.09 \pm 0.19$ & $-2.14 \pm 0.18$ & $-2.10 \pm 0.36$ &
\nodata & \nodata & \nodata \\
N$^+$/O$^+$ &
\nodata & $0.027 \pm 0.008$ & $0.029 \pm 0.007$ &
\nodata & \nodata & \nodata \\
log(N/O) &
\nodata & $-1.57 \pm 0.12$ & $-1.53 \pm 0.09$ &
\nodata & \nodata & \nodata \\
Ne$^{+2}$/O$^{+2}$ &
$0.251 \pm 0.034$ & $0.276 \pm 0.034$ & $0.327 \pm 0.054$ &
\nodata & \nodata & \nodata \\
log(Ne/O) &
$-0.600 \pm 0.058$ & $-0.558 \pm 0.053$ & $-0.485 \pm 0.070$ &
\nodata & \nodata & \nodata \\
\\
\tableline
Property & HM 17 & HM 19\tablenotemark{d} & HM 19\tablenotemark{d} 
& HM 21 \\
& (gr \#11) & (gr \#11) & (gr \#11) & (gr \#11) \\
\tableline
12$+$log(O/H) M91\tablenotemark{b} & 8.17 & 8.24 & 7.64 & 8.14 \\
12$+$log(O/H) P00\tablenotemark{c} & \nodata & 8.40 & 8.12 & 8.28 \\
\tableline
\vspace*{1mm}
\end{tabular}
\tablenotetext{a}{
HM~12: see Table~\ref{table_data3}. 
}
\tablenotetext{b}{
\cite{mcgaugh91} bright-line calibration.
}
\tablenotetext{c}{
\cite{pilyugin00} bright-line calibration.
}
\tablenotetext{d}{
HM~19: see Table~\ref{table_data4}. 
}
\tablecomments{
See Table~\ref{table_abund1} for comments.
}
\end{center}
\end{table}


\begin{table}
\small  
\tablenum{6}
\begin{center}
\renewcommand{\arraystretch}{1.1}
\caption{
Comparison with previous spectroscopic data.
\vspace*{3mm}
\label{table_compare}
}
\begin{tabular}{ccccccc}
\tableline \tableline
& \multicolumn{2}{c}{STM89\tablenotemark{a}} &
\multicolumn{2}{c}{HM95} &
\multicolumn{2}{c}{Present work} \\
Property & HM 2 & HM 9 & HM 7 & HM 9 & 
HM 7 & HM 9\tablenotemark{b} \\
\tableline
%
%
O/H ($\times 10^5$) &
\phs$5 \pm 3$ & \phs$6 \pm 3$ & 
\phs$5.2 \pm 0.8$ & \phs$6.5 \pm 1.1$ &
\phs$5.22 \pm 0.50$ & \phs$8.16 \pm 0.79$ \\
12$+$log(O/H) &
\phs$7.70 \pm 0.30$ & \phs$7.78 \pm 0.24$ & 
\phs$7.72 \pm 0.07$ & \phs$7.81 \pm 0.08$ &
\phs$7.72 \pm 0.04$ & \phs$7.91 \pm 0.04$ \\
log(N/O) &
\nodata & $-1.3 \pm 0.2$ & 
$-1.60 \pm 0.15$ & $-1.00 \pm 0.24$ &
$-1.46 \pm 0.05$ & $-1.55 \pm 0.08$ \\
log(Ne/O) &
\phn $-0.8 \pm 0.2$ \phn & \phn $-0.6 \pm 0.2$ \phn & 
\phn $-0.87 \pm 0.15$ \phn & \phn $-0.94 \pm 0.18$ \phn &
\phn $-0.594 \pm 0.046$ \phn & \phn $-0.562 \pm 0.035$ \phn \\
log(Ar/O) &
\nodata & $-2.4 \pm 0.2$ & 
$-2.37 \pm 0.14$ & $-2.22 \pm 0.18$ &
$-2.25 \pm 0.10$ & $-2.12 \pm 0.12$ \\
\tableline
\vspace*{1mm}
\end{tabular}
\tablenotetext{a}{
\cite{stm89} labelled their \hii\ regions ``\#1'' (HM~9) and
``\#2'' (HM~2).
}
\tablenotetext{b}{
Mean of three \othreea\ measurements; see Table~\ref{table_abund2}.
}
\tablerefs{
STM89 - \cite{stm89}, HM95 - \cite{hm95}.
}
\end{center}
\end{table}

\end{document}